\documentclass[prx,twocolumn,showpacs,superscriptaddress,preprintnumbers,amssymb]{revtex4-2}
\usepackage{graphicx}
\usepackage{latexsym}
\usepackage{amsmath}
\usepackage{mathtools}
\usepackage{amsfonts}
\usepackage{upgreek}
\usepackage{bm}
\usepackage{multirow}
\usepackage{enumitem}
\usepackage{color}
\usepackage[colorlinks, citecolor=blue]{hyperref}
\usepackage{physics}
\usepackage{tcolorbox}
\usepackage{manfnt}
\usepackage{relsize}

\newcommand{\beq}{\begin{equation}}
\newcommand{\eeq}{\end{equation}}
\newcommand{\beqn}{\begin{eqnarray}}
\newcommand{\eeqn}{\end{eqnarray}}

\newcommand{\vphi}{\varphi}

\newcommand{\cO}{ {\cal O} }

\newcommand{\secref}[1]{Sec.\,\ref{#1}}
\newcommand{\appref}[1]{Appendix.\,\ref{#1}}
\newcommand{\eqnref}[1]{Eq.\,\eqref{#1}}

\newcommand{\figref}[1]{Fig.\,\ref{#1}}

\newcommand{\rAngle}{\rangle \hspace{-2pt} \rangle }
\newcommand{\lAngle}{\langle \hspace{-2pt} \langle }

\begin{document}

\title{Holographically Emergent Gauge Theory in Symmetric Quantum Circuits} 

\author{Akash Vijay}
\affiliation{Department of Physics and Anthony J. Leggett Institute of Condensed Matter Theory, University of Illinois at Urbana-Champaign, Urbana, Illinois 61801, USA}

\author{Jong Yeon Lee}
\affiliation{Department of Physics and Anthony J. Leggett Institute of Condensed Matter Theory, University of Illinois at Urbana-Champaign, Urbana, Illinois 61801, USA}

\date{\today}
\begin{abstract}
We develop a novel holographic framework to study dynamical phases in random quantum circuits with a global symmetry $G$. Viewing the circuit as a tensor network, we decompose it into two parts: a symmetric layer, which defines an emergent gauge wavefunction in one higher dimension, and a non-symmetric layer, composed of random multiplicity tensors.
For $G\,{=}\,\mathbb{Z}_N$ symmetric circuits consisting of local unitary gates interspersed with local symmetric noise channels, averaging over the non-symmetric layer yields a \emph{dynamically generated} noisy $\mathbb{Z}_{N}$ surface code. Distinct global charge sectors of the spin chain map to different topological sectors of the bulk gauge theory. This allows us to interpret $\mathbb{Z}_{N}$ symmetric circuits in the volume-law phase as quantum error-correcting codes with a distinguished set of logical spin states that inherit the topological protection of the bulk code. By establishing equality of bulk and boundary coherent information, we show that quantum information encoded in these logical states is maximally protected against symmetric noise up to a finite threshold. We further study weakly monitored $\mathbb{Z}_{N}$ symmetric circuits which exhibit a charge-sharpening transition. We show that the point at which the observer gains classical information about the global charge coincides with the point at which measurements destroy the underlying quantum information encoded in the bulk surface code. This also allows for a natural interpretation of the sharpening transition as a confinement transition in the gauge theory. For $N\,{\leq}\,4$, weak measurements drive a single transition from a charge-fuzzy phase with exponential sharpening time $t_{\#}\sim e^{L}$ to a charge-sharp phase with $t_{\#}\sim \mathcal{O}(1)$. On the other hand, for $N>4$, the circuit can enter an intermediate phase with a linear sharpening time $t_{\#}\sim \mathcal{O}(L)$. In this regime, the bulk gauge theory realizes a Coulomb phase with emergent gapless photons.
\end{abstract}
\maketitle

\section{Introduction}

Measurement-induced phase transitions (MIPTs) are now recognized as universal phenomena which appear in open quantum systems undergoing local scrambling and measurement dynamics~\cite{MIPT2018,MIPT2019,MIPT2019_2,MIPT_Notes,MIPT_Percolation,MIPT_weakmeasurements,MIPT_FreeFermion,MIPT_Ising,Purification_Transition,MIPT_ErrorCorrection,MIPT_All_to_All,MIPT_Critical_Measurement_Rate,MIPT_CriticalExponents,MIPT_CriticalityCentralCharge,MIPT_CriticalityCFT,MIPT_CriticalityProbes,RG_for_MIPTS}. 
As the measurement rate $p$ is increased past a critical value $p_c$, the steady-state entanglement entropy of subregions exhibits a sharp transition from a volume law to an area law. Equivalently, MIPTs can be viewed as \textit{purification transitions} whereby, once $p > p_c$, an initially mixed state purifies on a timescale $t_p \sim \mathcal{O}(1)$ that is independent of system size~\cite{Purification_Transition}. They may also be interpreted as \textit{error-correction transitions}, in which sufficiently frequent measurements overwhelm the scrambling dynamics and prevent quantum information from being hidden from the measuring observer~\cite{Purification_Transition,MIPT_ErrorCorrection}.

Incorporating symmetries into the dynamics enriches the phase diagram and qualitatively modifies both operator spreading and entanglement growth~\cite{OperatorSpreading1,OperatorSpreading2,Entanglement_Dynamics_U1_1,Entanglement_Dynamics_U1_2,Proof_Diffusive}. For instance, in 1D U(1)-symmetric circuits with local charge measurements, the volume-law phase splits into two phases, charge sharp and charge fuzzy, at a second critical measurement rate \(p_{\#} < p_c\)~\cite{U1_ChargeSharpening1,U1_ChargeSharpening2,Field_Theory_ChargeSharpening,No_Sharpening_Area_Law_Phase,Phases_Symmetry_Quantum_Circuits,Field_Theory_ChargeSharpening,SU2_Sharpening,Sharpening_hydrodynamics,Charge_Sharpening_Trees,Bayesian_CriticalPoints}. This \emph{charge-sharpening transition} is diagnosed by the timescale \(t_{\#}\) needed to infer the global charge from local measurement records: for \(p < p_{\#}\), \(t_{\#} \sim L\), while for \(p > p_{\#}\), \(t_{\#} \sim \mathcal{O}(1)\), even though identifying the microscopic state still requires exponentially long times.

Measurements introduce intrinsic stochasticity into the dynamics, with each sequence of outcomes generating a distinct final state or ``trajectory''. Each trajectory occurs with a probability determined by the Born rule. Measurement-induced phase transitions arise from qualitative changes in the ensemble of such trajectories; they are {dynamical phase transitions}.  
Developing a complete classification of such orders in monitored, out-of-equilibrium quantum systems remains a central challenge.

\begin{figure*}
    \includegraphics[width=0.98\textwidth]{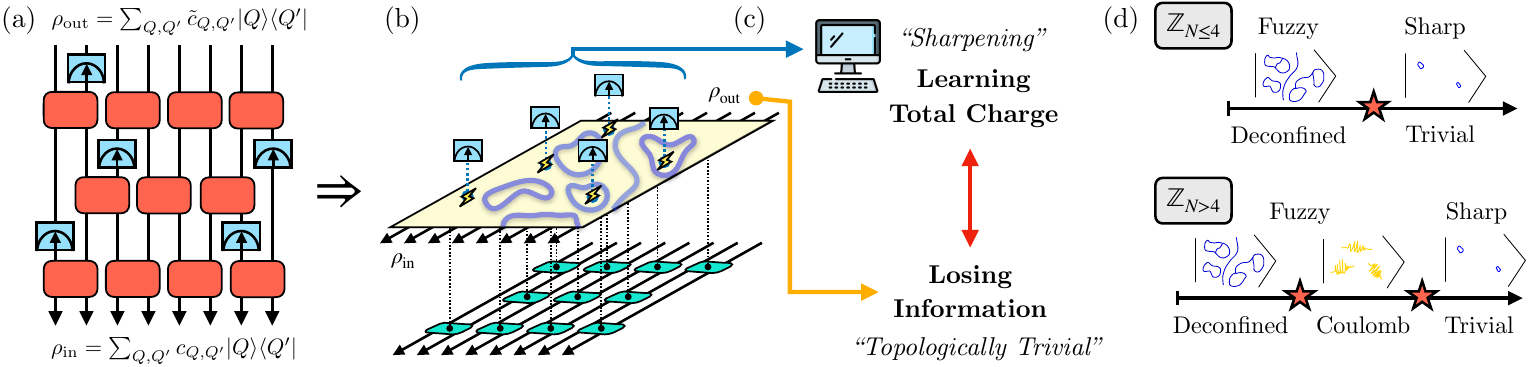}
    \caption{\label{fig:Two_Layer_Decomposition}  {\bf Summary.} 
    {\bf (a) Monitored Symmetric Circuit.} 
    The circuit consists of two parts: local symmetric gates and local charge measurements. 
    One can map a symmetric monitored random circuit into a tensor network.
    {\bf (b) Two-layer tensor network.}
    The top layer of the network is constructed using the Clebsch-Gordon coefficients of the group and realizes a gauge theoretical wavefunction.
    The bottom layer is comprised of random multiplicity tensors. 
    {\bf (c) Sharpening transition.} 
    Charge sharpening occurs when local charge measurements extract the global charge of the spin chain. At the same time, quantum information encoded in the bulk gauge wavefunction is destroyed. Therefore, fuzzy (sharp) phase corresponds to information-protected (lost) phase in the gauge theory.
    {\bf (d) Phase diagram.} For $\mathbb{Z}_{N\leq4}$, there exists only two phases: truly fuzzy (deconfined) and sharp (topologically trivial/confined), where sharpening time scales as $\exp(L)$ and $\cO(1)$, respectively. However, for $\mathbb{Z}_{N>4}$ an intermediate (Coulomb) phase opens up, where the sharpening time scales as $\cO(L)$. 
    }
\end{figure*}
Here we propose a novel framework to understand the dynamical phases arising in random symmetric circuits through the lens of a \textit{holographically emergent gauge theory}. This framework is general, enabling the study of such phases in circuits with arbitrary Abelian or non-Abelian symmetries. Leveraging powerful tensor network techniques~\cite{TNS_GlobalSymmmetry_Vidal,TNS_GlobalSymmmetry_U1,TNS_GlobalSymmmetry_SU2, qi2022emergentbulkgaugefield}, we decompose symmetric random circuits into a two-layered tensor network: a symmetric layer describing propagation of charged degrees of freedom and a non-symmetric layer describing propagation of uncharged ones (see ~\figref{fig:Two_Layer_Decomposition}). We show that for a $1$ dimensional quantum circuit endowed with a global $G$-symmetry, this symmetric layer constitutes a $2$ dimensional $G$-gauge wavefunction (or more generally a $G$-\textit{spin network state}~\cite{Spin_Networks_Penrose,Spin_Networks_Baez,Spin_Networks_QG}).

Controlled calculations are made possible when the uncharged layer has large bond dimension. 
For random circuits composed of unitary gates, averaging over the non-symmetric layer in the large bond dimension limit  produces a bulk gauge state given by a coherent superposition of all allowed string-net configurations. 
When $G = \mathbb{Z}_{N}$, we obtain precisely the deconfined fixed point state of $\mathbb{Z}_{N}$ gauge theory. In this case, the global symmetry of the spin chain maps to a 1-form symmetry in the gauge theory, with different global charge sectors mapping to distinct topological sectors.

We further extend this bulk-boundary correspondence to the case of spin chains evolving under $\mathbb{Z}_{N}$ symmetric unitary gates and subject to $\mathbb{Z}_{N}$ symmetric local noise channels at each timestep. In this setting, we show that the bulk gauge theory realizes a \emph{dynamically generated} $\mathbb{Z}_{N}$ surface code~\cite{Kitaev_Topological_Error_Correction, Surface_Codes_Definition,Dennis_2002_Error_Threshold}, with noise acting on every link of the code. This observation naturally identifies a set of logical states for the spin chain, one within each global charge sector, which inherit the topological protection of the bulk surface code. Consequently, in this phase the symmetric circuit functions as a stable quantum memory, encoding one qudit of quantum information that remains robust against sufficiently weak symmetry-preserving noise. We make this correspondence precise by showing that the coherent information of the spin chain evolving under the circuit+noise channel, after averaging over the uncharged layer, is exactly equal to the coherent information of the bulk topological code subject to the same noise channel. This establishes a sharp equivalence between the dynamical phases of the noisy symmetric quantum circuit in its volume-law phase and the mixed-state phases of a noisy topological code.

We further consider $\mathbb{Z}_{N}$ symmetric circuits subject to weak (imperfect) charge measurements of every spin at each timestep. In the bulk gauge description, this yields a $\mathbb{Z}_{N}$ surface code subject to weak measurements of every link of the code. Charge sharpening corresponds to a measurement-induced decodability transition in the bulk, whereby logical information is destroyed by the measuring observer. We find that the point at which the observer gains classical information about the global charge coincides with the point at which measurements destroy the underlying quantum information encoded in the logical spin states. Furthermore, we show that this phenomenon admits a natural interpretation in the bulk as a confinement transition~\cite{Separability_Trannsition_FiniteT,Separability_Transitions_Topological_Codes}. In the charge-fuzzy phase, the post-measurement ensemble of gauge wavefunctions are deconfined leading to an exponentially large sharpening timescale $(t_{\#}\sim e^{L})$. However, once the system crosses the sharpening threshold, the gauge states become flux condensates resulting in a rapid sharpening timescale $(t_{\#}\sim \mathcal{O}(1))$. When $N>4$, we find that the circuit can enter an intermediate phase with a linear sharpening timescale $(t_{\#}\sim L)$, mirroring the behavior found in the U(1) symmetric circuits. In this regime, the bulk gauge theory enters a Coulombic phase featuring emergent gapless photons and power-law correlations. Although our formalism is fully general, in this work we focus primarily on $\mathbb{Z}_{N}$-symmetric circuits. The rich structure of phases arising in non-abelian symmetric circuits will be explored in a forthcoming work.

\section{Emergent Gauge State from random symmetric circuits}

In this section, we explain how a gauge-theory wavefunction naturally emerges from quantum circuits with an on-site global symmetry \(G\)~\cite{qi2022emergentbulkgaugefield,TNS_GlobalSymmmetry_Vidal,TNS_GlobalSymmmetry_U1,TNS_GlobalSymmmetry_SU2,Background_Independent_Tensor_Networks}.

\begin{figure}
\includegraphics[width=0.95\linewidth]{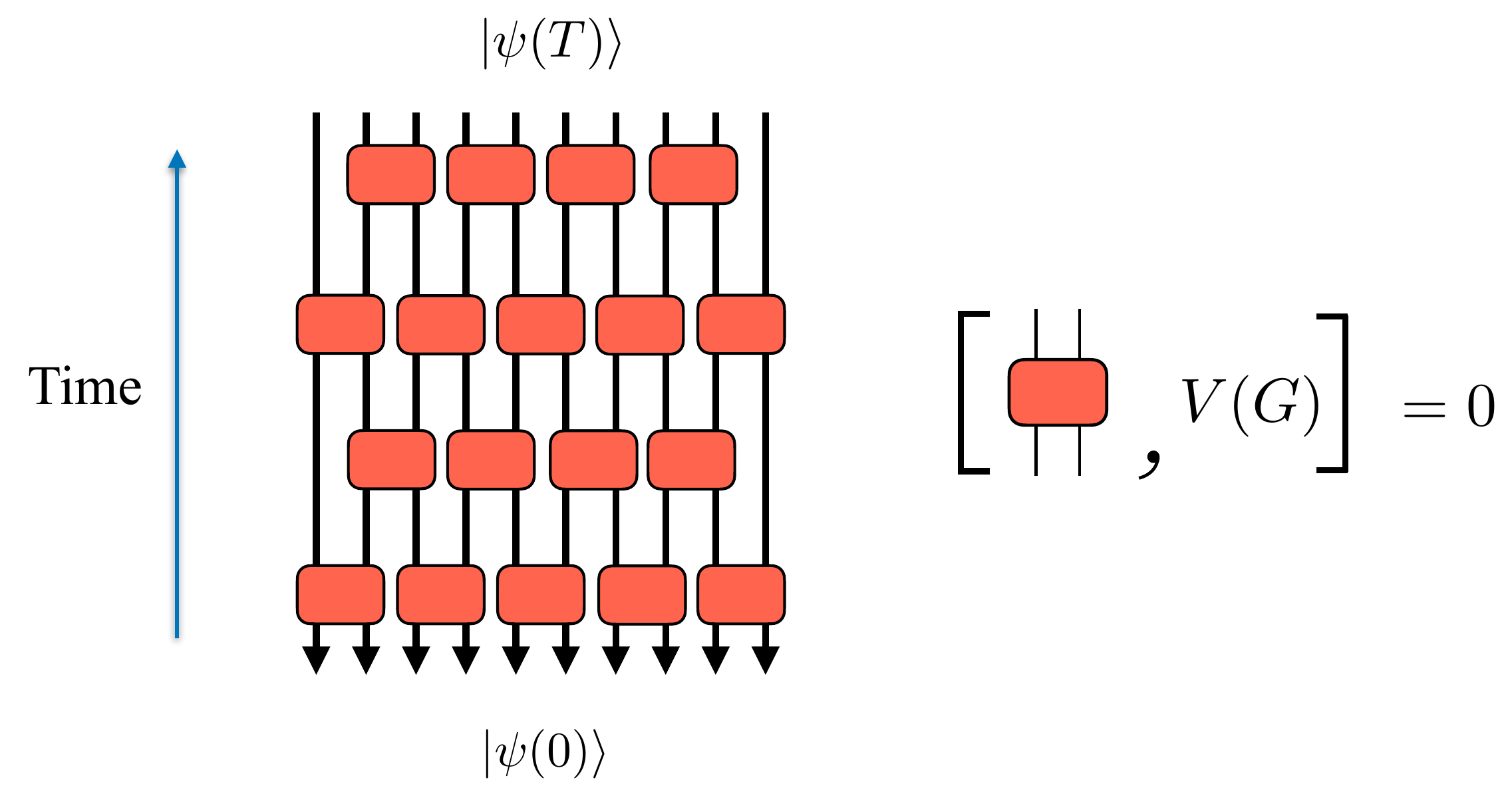}
\caption{\label{fig:Symmetric_Circuit}  {\bf Random Symmetric Circuits:} We consider a random symmetric circuit with bricklayer geometry as depicted. The symmetry group $G$ acts by means of some unitary representation \(\mathcal{V}(\cdot):G \rightarrow \mathcal{U}(\mathcal{H}_{\text{global}})\). Each gate is individually chosen to commute with the group action but is otherwise random.} 
\end{figure}

We consider a spin chain of length \(L\), with a local Hilbert space \(\mathcal{H}\) of dimension \(q\). The global Hilbert space is \(\mathcal{H}_{\text{global}} = \mathcal{H}^{\otimes L}\) of dimension \(q^{L}\). 
Starting from an initial state \(|\psi_{i}\rangle \in \mathcal{H}_{\text{global}}\), we evolve with a (possibly non-unitary) brickwork circuit operator \(\mathcal{U}_{\text{circuit}}\) built from $k$-local gates (see \figref{fig:Symmetric_Circuit} where $k = 2$). 

We can transform each $k$-local gate, \(U:\mathcal{H}^{\otimes k}\rightarrow \mathcal{H}^{\otimes k}\), to a (vertex) state, \(|U\rangle\in\mathcal{H}^{\otimes k}\otimes \mathcal{H}^{*\otimes k}\), by vectorizing its singular value decomposition~\cite{CHOI1975, JAMIOLKOWSKI1972}:
\begin{align}
         U = \sum_{i} \lambda_{i}|l_{i}\rangle\langle r_{i}| \rightarrow |U\rangle = \sum_{i} \lambda_{i}|l_{i}\rangle|r^{*}_{i}\rangle.
\end{align}
Here $|r^{*}_{i}\rangle$ denotes the complex-conjugated state. This operation is basis dependent. In the charge basis introduced below, it admits a natural interpretation as charge conjugation.

Different vertex states are connected to one another by projecting them along maximally entangled EPR pairs, \(|I\rangle = \sum_{i}|i\rangle|i^{*}\rangle\). The latter just corresponds to the vectorization of the identity operator. 
In this notation, the final state of the spin chain \(|\psi_{f} \rangle = \mathcal{U}_{\text{circuit}}|\psi_{i} \rangle\), can be written as 
\begin{align} \label{eq:temp2}
    |\psi_{f} \rangle =\Big[ \langle \psi_{i}^{*}| 
    \otimes \langle \bm{I} | \Big] \cdot  | \bm{U} \rangle,
\end{align}
where \(| \bm{U} \rangle := \otimes_v |U_v \rangle\) is the tensor product of all gate states, \(| \bm{I} \rangle := \otimes_l |I_l \rangle\) is the tensor product of all EPR pairs on links, and \(|\psi^{*}_{i} \rangle\) is the complex-conjugated initial state. 


\subsection{Charge Basis}

Symmetry is implemented by a unitary representation \(\mathcal{V}(\cdot): G \rightarrow \mathcal{U}(\mathcal{H}_{\text{global}})\) acting locally on the spin chain as \(\mathcal{V}(g) =  V(g)^{\otimes L}\). Here \({V}(\cdot): G \rightarrow \mathcal{U}(\mathcal{H})\) is itself a unitary representation on the local Hilbert space $\mathcal{H}$. A gate $U$ preserves this symmetry if
\begin{align}\label{eq:symmetry_gate}
    [U,\mathcal{V}(g)] = 0,
\end{align}
for all \( g \in G\). Since any finite-dimensional representation decomposes into a sum of irreducible representations (irreps), the local Hilbert space can be written as
\begin{align}
    \mathcal{H} =  \bigoplus_{j} n_{j}\mathcal{R}_{j}.
\end{align}
Here \(\mathcal{R}_{j}\) is a $G$-irrep labeled by $j$ and $n_{j}$ is its multiplicity. The charge basis is given by
\begin{align} \label{eq:basis}
    |{j,\mu_{j},a_{j}}\rangle,
\end{align} 
where $j$ labels the irrep (``charge"), $\mu_{j}$ denotes a state within irrep $j$, and $a_{j}$ is the multiplicity index associated with irrep $j$. Whenever possible, we omit the subscripts on $\mu_{j}$ and $a_{j}$ to simplify notation. Note that the symmetry operators $V(g)$ are block-diagonal in the irrep labels and act trivially on the multiplicity labels,
\begin{align} \label{eq:symmetry_operator}
    {V}(g) = \bigoplus_{j} V^{j}(g) \otimes \mathbb{I}_{n_{j}}.
\end{align}
\(\mathbb{I}_{n_{j}}\) is the identity operator on the $n_j$-dimensional multiplicity space of irrep $j$. Complex conjugation is defined in this basis through the action \(|j,\mu_{j},a_{j}\rangle \rightarrow |{j^{*},\mu_{j},a_{j}}\rangle\), where $j^{*}$ denotes the complex conjugate of irrep $j$.

\subsection{Symmetry Decomposition of Vertex Tensors} 
\begin{figure}
    \includegraphics[width=0.95\linewidth]{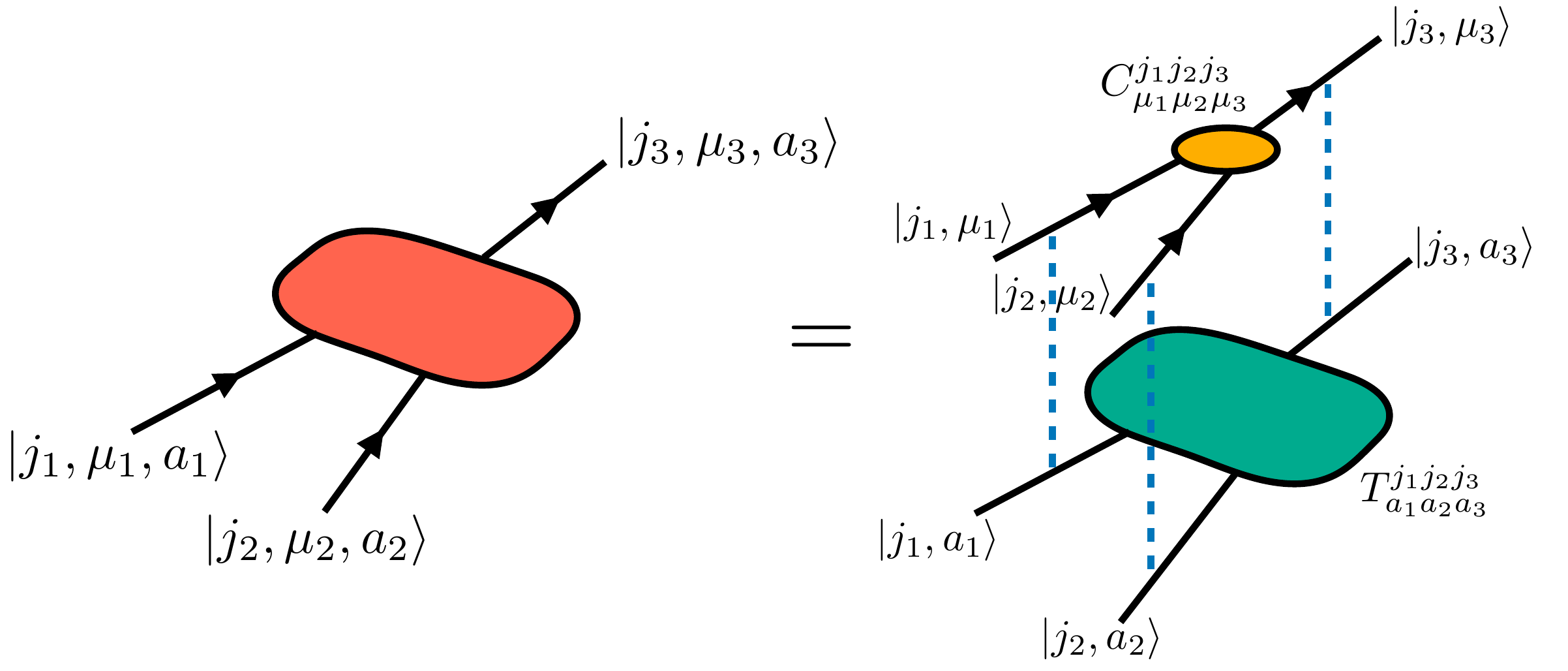}
    \caption{\label{fig:Symmetry_Decomposition3}  {\bf Symmetry Decomposition I}. Here we depict the symmetry decomposition of a trivalent vertex with two incoming legs and one outgoing leg.}
\end{figure}

\begin{figure}
    \includegraphics[width=0.95\linewidth]{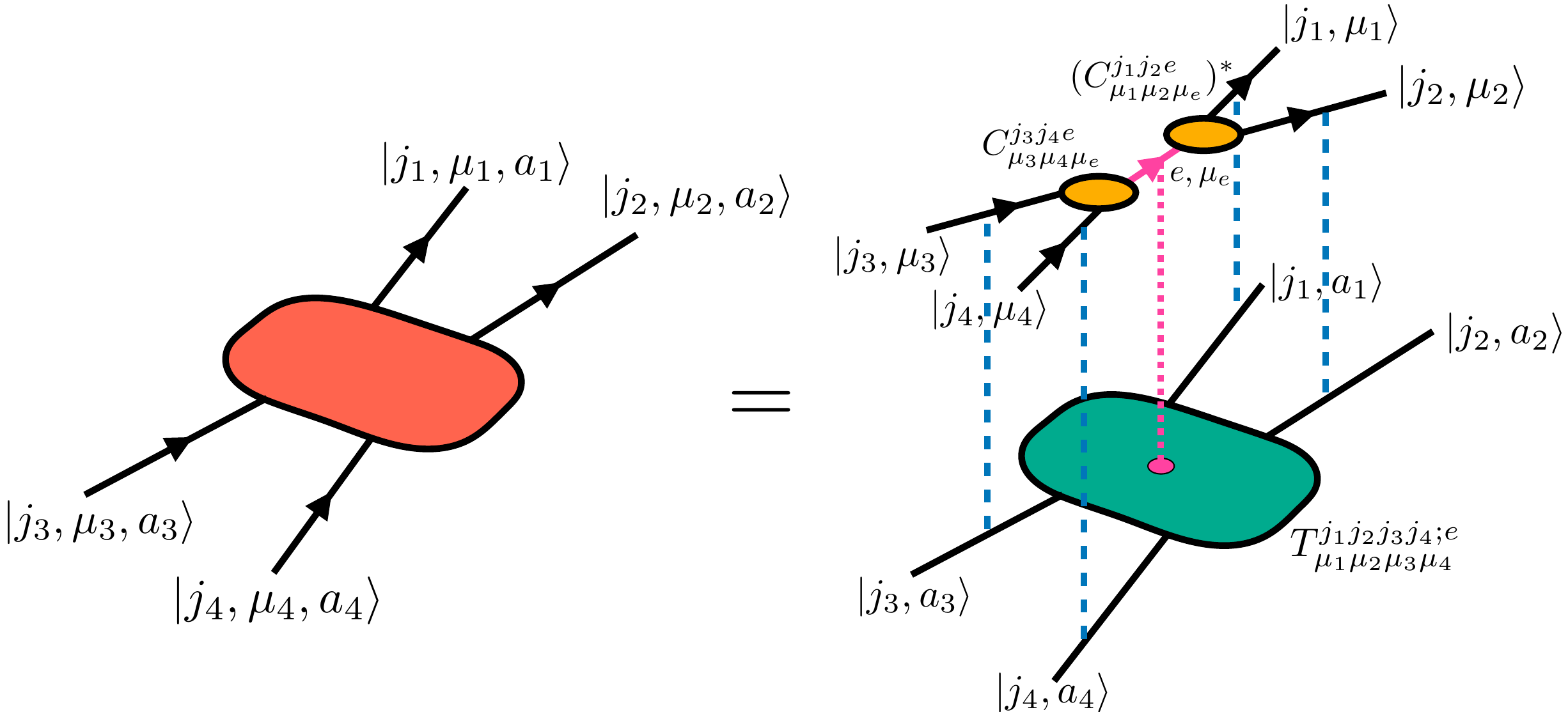}
    \caption{\label{fig:Symmetry_Decomposition4}  {\bf Symmetry Decomposition II}. Here we depict the symmetry decomposition of a four-legged vertex with two incoming legs and two outgoing legs.}
\end{figure}

Under a generic symmetry transformation \(g \in G\), each of the \(k\)-``ket'' legs of \(|{U}\rangle\) transforms under \(V(g)\) whereas each of the \(k\)-``bra'' legs transforms under \(V(g)^{*}\) (since they carry complex conjugated irrep labels). Thus, the full vertex state transforms as
\begin{align}
    |U\rangle \rightarrow V(g)^{\otimes k} \otimes V(g)^{*{\otimes k}} |U\rangle.
\end{align}
We denote the ``bra" legs with incoming arrows and the ``ket" legs with outgoing arrows (see \figref{fig:Symmetry_Decomposition3} and \figref{fig:Symmetry_Decomposition4}). Symmetry of the local gate (\eqnref{eq:symmetry_gate}) requires that the corresponding vectorized state \(|{U}\rangle\) be a group singlet under the action of \(G\).

We can expand \(\ket{U}\) in the basis given in \eqnref{eq:basis}. Introducing a joint index \(J = (j,\mu_j,a_j)\), with \(\mathbf{J} = (J_{1},\dots,J_{2k})\), we write
\begin{align}
    |U\rangle = \sum_{\mathbf{J}} A^{\mathbf{J}} \otimes_{i=1}^{2k} |J_{i}\rangle,
\end{align}
where \(A^{\mathbf{J}}\) is a \(G\)-invariant tensor with one index per leg. Representation theory then constrains the structure of such tensors~\cite{qi2022emergentbulkgaugefield,TNS_GlobalSymmmetry_Vidal,TNS_GlobalSymmmetry_U1,TNS_GlobalSymmmetry_SU2}. 

For instance, consider a two-legged tensor \(A^{J_{1}J_{2}}\) with one incoming and one outgoing leg. The invariance condition reads
\begin{align}
    A^{J_{1}J_{2}} = \sum_{J_{1}',J_{2}'} V(g)^{J_{1}J_{1}'} A^{J_{1}'J_{2}'} [V(g)^{\dagger}]^{J_{2}'J_{2}},
\end{align}
for all \(g \in G\). By Schur's lemma, such an invariant tensor must be of the form,
\begin{align}
    A^{J_{1}J_{2}} = \delta^{j_{1}j_{2}} \delta_{\mu_{1}\mu_{2}} T^{j_{1}j_{2}}_{a_{1}a_{2}},
\end{align}
where \(T^{j_{1}j_{2}}_{a_{1}a_{2}}\) is an unconstrained multiplicity tensor. The generalization to the case in which both legs are incoming or both legs are outgoing is obtained by conjugating the corresponding irrep labels.

Consider now a three-legged tensor \(A^{J_{1}J_{2}J_{3}}\) with, say, two incoming and one outgoing leg. The irreps obey a fusion algebra
\begin{align}
    \mathcal{R}_{i} \otimes \mathcal{R}_{j} = \bigoplus_{k} N^{k}_{ij}\mathcal{R}_{k}
\end{align}
where \(N_{ij}^{k}\) is the multiplicity of the channel \(i \times j \to k\). For notational simplicity, we assume the group is multiplicity-free in fusion, \(N_{ij}^{k} \leq 1\). The Clebsch-Gordan coefficients,
\begin{align}
    C^{j_{1}j_{2}j_{3}}_{\mu_{1}\mu_{2}\mu_{3}} = \braket{j_{3},\mu_{3}}{j_{1},\mu_{1}; j_{2},\mu_{2}},
\end{align}
are invariant tensors of the group~\footnote{This is the complex conjugate of how these coefficients are usually defined which would correspond to the case of two incoming legs and one outgoing leg}. By the Wigner-Eckart theorem, any three-legged invariant tensor can be decomposed as
\begin{align}
    A^{J_{1}J_{2}J_{3}} = C^{j_{1}j_{2}j_{3}}_{\mu_{1}\mu_{2}\mu_{3}} T^{j_{1}j_{2}j_{3}}_{a_{1}a_{2}a_{3}},
\end{align}
where \(T^{j_{1}j_{2}j_{3}}_{a_{1}a_{2}a_{3}}\) is an unconstrained multiplicity tensor. In other words, symmetry splits the tensor into a fixed \emph{charged} part $C$ (the intertwiner) and an unconstrained \emph{multiplicity} part $T$, see \figref{fig:Symmetry_Decomposition3}.

\begin{figure}
    \includegraphics[width=0.95\linewidth]{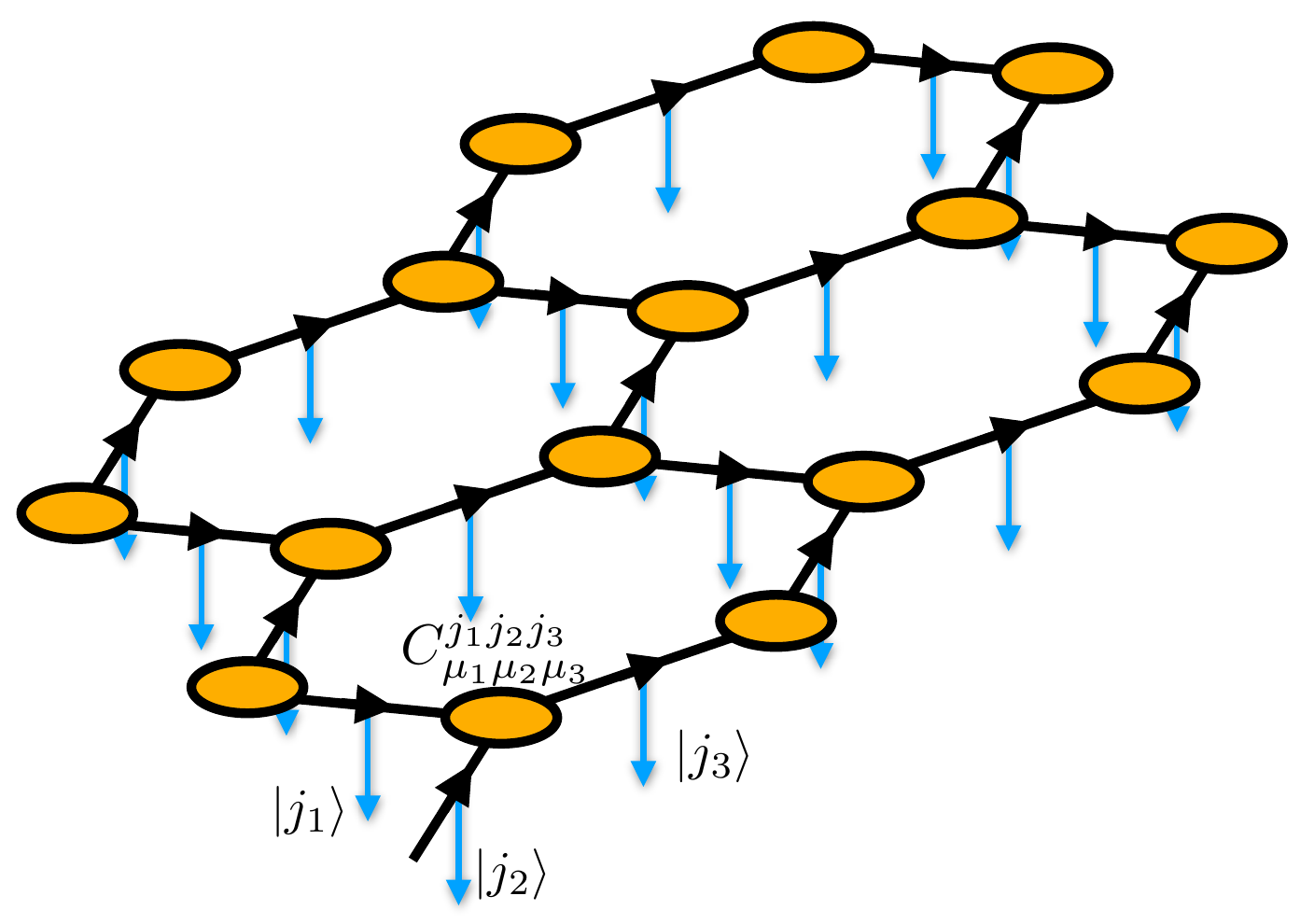}
    \caption{\label{fig:Gauge_Wavefunction}  {\bf Gauge wavefunction in the charged layer.} Decomposing four-legged symmetric vertices in terms of trivalent Clebsch-Gordan tensors transforms the square lattice into a hexagonal lattice with trivalent vertices. Link degrees of freedom carry irrep labels \(j_{l}\), while internal indices \(\mu\) are summed over. Clebsch-Gordan tensors act as intertwiners enforcing an exact Gauss law. The resulting spin-network state can be interpreted as a state in a lattice gauge theory. }
\end{figure}

A similar structure holds for four-legged tensors. Consider a tensor \(A^{J_{1}J_{2}J_{3}J_{4}}\) with two outgoing legs labeled by \(j_{1},j_{2}\) and two incoming legs labeled by \(j_{3},j_{4}\). If \(j_{1} \otimes j_{2}\) and \(j_{3} \otimes j_{4}\) can both fuse to irrep \(e\), an invariant tensor can be built as
\begin{align} \label{eq:temp1}
    B^{j_{1}j_{2}j_{3}j_{4}; e }_{\mu_{1}\mu_{2}\mu_{3}\mu_{4}} 
    = \sum_{\mu_{e}} \big(C^{j_{1}j_{2}e}_{\mu_{1}\mu_{2}\mu_{e}}\big)^{*} C^{j_{3}j_{4}e}_{\mu_{3}\mu_{4}\mu_{e}}.
\end{align}
There is one such invariant tensor for each allowed intermediate charge \(e\). A generic four-legged invariant tensor thus decomposes as
\begin{align}
    A^{J_{1}J_{2}J_{3}J_{4}} 
    = \sum_{e} B^{j_{1}j_{2}j_{3}j_{4}; e }_{\mu_{1}\mu_{2}\mu_{3}\mu_{4}} 
    T^{j_{1}j_{2}j_{3}j_{4}; e}_{a_{1}a_{2}a_{3}a_{4}},
\end{align}
where \(T^{j_{1}j_{2}j_{3}j_{4}; e}_{a_{1}a_{2}a_{3}a_{4}}\) is again an unconstrained multiplicity tensor, see \figref{fig:Symmetry_Decomposition4}. We can interpret \eqnref{eq:temp1} as a way to split the four-legged vertex into a pair of three-legged vertices connected by an internal link which itself carries charge and state labels. The result of this splitting is to transform the square lattice of the charged layer into a hexagonal lattice with Clebsch-Gordon tensors sitting at each trivalent vertex (\figref{fig:Gauge_Wavefunction}).

\begin{figure}
    \includegraphics[width=0.95\linewidth]{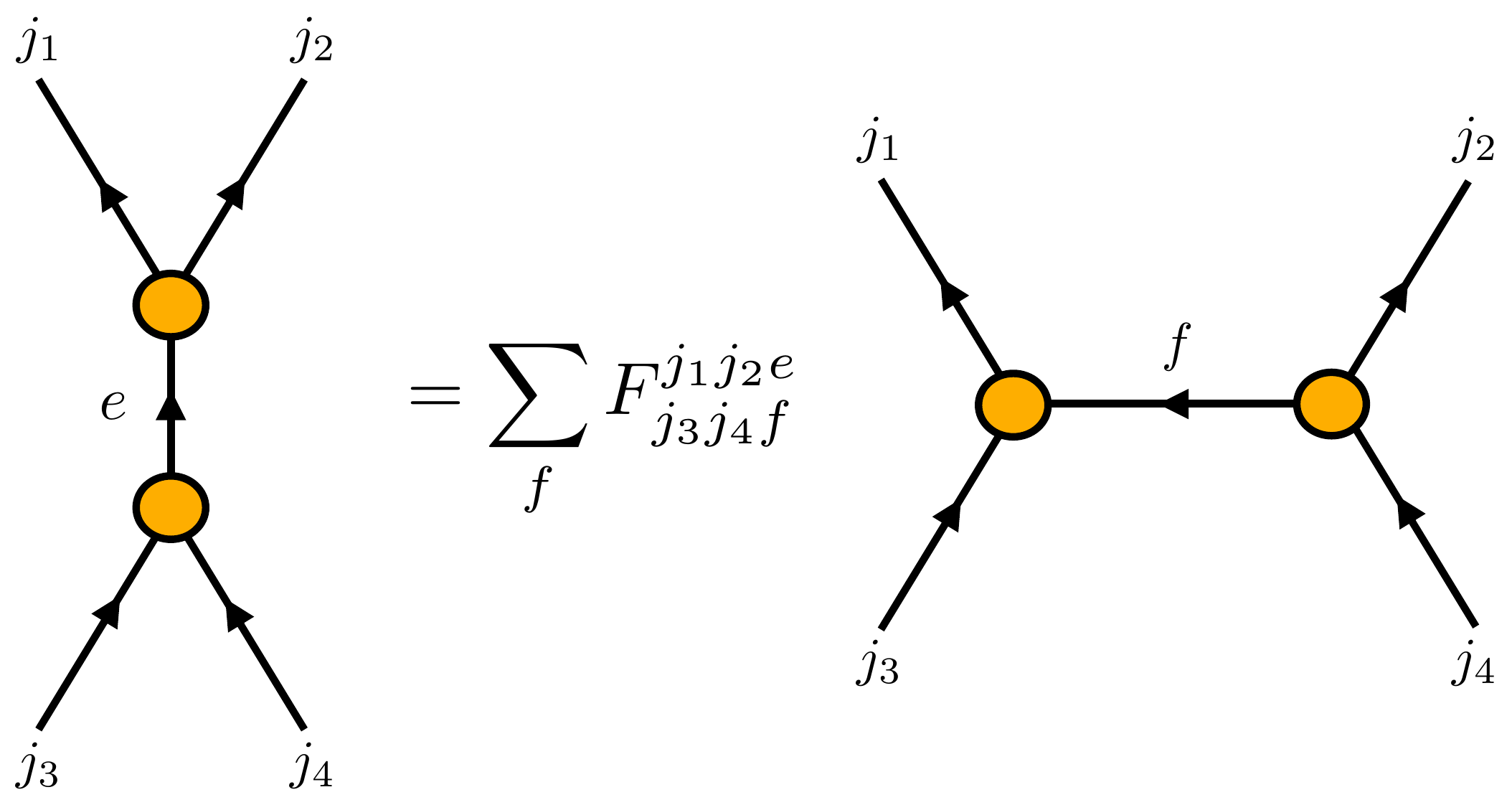}
    \caption{\label{fig:Fmove}  {\bf $F$-move.} Different trivalent decompositions of a four-legged invariant tensor are related by the F-symbol.}
\end{figure}

We can equivalently build invariant tensors by choosing an alternative fusion channel, in which $j_{2}^{*}\otimes j_{4}$ fuses to an intermediate irrep $f$ and $j_{3}\otimes f$ fuses to $j_{1}$. This yields
\begin{align}
    \tilde{B}^{j_{1}j_{2}j_{3}j_{4}; f }_{\mu_{1}\mu_{2}\mu_{3}\mu_{4}} 
    = \sum_{\mu_{f}} C^{j_{3}fj_{1}}_{\mu_{3}\mu_{f}\mu_{1}} C^{j_{2}^{*}j_{4}f}_{\mu_{2}\mu_{4}\mu_{f}}.
\end{align}
Invariant tensors obtained from different trivalent decompositions of a four-valent vertex are related by linear $F$-move transformations,
\begin{align}
    B^{j_{1}j_{2}j_{3}j_{4}; e} = \sum_{f} F^{j_{1}j_{2}e}_{j_{3}j_{4}f} \tilde{B}^{j_{1}j_{2},j_{3},j_{4}; f},
\end{align}
where \(F\) is the F-symbol of the underlying unitary fusion category, see \figref{fig:Fmove}. These F-symbols obey the pentagon equation~\cite{6j_Pentagon_Equation}. The resulting structure is that of a unitary modular tensor category, familiar from the description of bosonic topologically ordered phases.

For tensors with more legs, one iteratively decomposes multi-legged vertices in terms of trivalent intertwiners, summing over the internal charges on all intermediate links. Different trivalent graphs are related by sequences of $F$-moves. We refer to this insensitivity to the underlying trivalent graph as \emph{background independence}~\cite{Background_Independent_Tensor_Networks}.

The Abelian case is particularly simple. All irreps are one-dimensional and the fusion of any two irreps produces a single irrep. The Clebsch-Gordan tensors reduce to Kronecker delta functions enforcing charge conservation. Thus, the symmetry constraint for a \(2k\)-legged tensor takes the form
\begin{align}
    A^{J_{1},\dots,J_{2k}} 
    = B^{j_{1},\dots,j_{2k}} T^{j_{1},\dots,j_{2k}}_{a_{1},\dots,a_{2k}},
\end{align}
where \(B^{j_{1},\dots,j_{2k}} = 1\) if the Abelian charge-selection rule is satisfied at the vertex and zero otherwise. There is no need to explicitly decompose into trivalent tensors in this case.

Collecting these observations, we find that each symmetric vertex tensor can be decomposed as 
\begin{align}
    A^{\mathbf{J}} = \sum_{\mathbf{e}}B^{\mathbf{j};\mathbf{e}}_{\boldsymbol{\mu}} T^{\mathbf{j};\mathbf{e}}_{\boldsymbol{a}},
\end{align}
where \(B\) encodes the Clebsch-Gordan data whereas \(T\) is an unconstrained multiplicity tensor. \(\mathbf{e}\) collectively referes to all intermediate charges. We refer to \(B\) as the \emph{charged tensor} and \(T\) as the \emph{multiplicity tensor}.
The full tensor network is obtained by contracting vertex states \(|{U}\rangle\) with EPR pairs. In the symmetry basis, each EPR pair takes the form $|{I}\rangle = \sum_{j,\mu,a} \ket{j,\mu,a} \ket{j^{*},\mu,a}$.


\subsection{Constructing the Gauge Wavefunction}

We now show that the charged layer, obtained by contracting Clebsch-Gordan tensors across the circuit, defines a gauge-theory wavefunction \(\ket{\Phi_{g}}\). For concreteness, we assume that the charged layer has been decomposed into a trivalent graph, with a Clebsch-Gordan tensor \(C_{v}\) assigned to each vertex \(v\).

We first fix an assignment of irrep labels \(\mathbf{j}\) and contract all the \(\boldsymbol{\mu}\)-indices. Contractions of \(\boldsymbol{\mu}\)-indices between neighboring intertwiners are nonzero only if the irrep labels on the two ends of each oriented link are conjugate to one another. Consequently, each link can be regarded as carrying a single charge label \(j_{l}\). After tracing over the internal state labels \(\boldsymbol{\mu}\), we obtain a wavefunction \(\Phi_{g}^\mathbf{j}\) defined over the configuration space of link charges. Explicitly, we write
\begin{align}\label{eq:gauge_wavefunction}
    \ket{\Phi_{g}}
    = \sum_{\{ j_{l} \}} 
    \operatorname{Tr}_{\boldsymbol{\mu}}
    \bigg[ \prod_{v} (C_{v})^{j_{1}j_{2}j_{3}}_{\mu_{1}\mu_{2}\mu_{3}} \bigg]
    \bigotimes_{l} \ket{j_{l}},
\end{align}
where the trace contracts \(\boldsymbol{\mu}\) indices on neighboring tensors, and the product runs over vertices of the trivalent graph. In analogy with PEPS, the link charges \(j_{l}\) play the role of ``physical'' degrees of freedom, while the \(\mu\)-indices are virtual.

Next we interpret the charge (irrep) label on each link as an electric flux (string) label of a gauge theory:
\begin{align}
    \text{Charges on links} \longleftrightarrow \text{Electric flux lines}.
\end{align}
The Clebsch-Gordan tensor \(C_{v}\) describes the fusion of flux lines meeting at vertex $v$. Since it is a group singlet, these states can be isometrically related to states of a lattice gauge theory with an exact Gauss law. 

To see this explicitly, recall that in a $G$-gauge theory, where $G$ is a finite (or compact Lie) group, the local Hilbert space on an oriented link $l$ is the set of square-integrable functions on $G$, $\mathcal{H}_{l} = \mathcal{L}^{2}(G)$, with a countable basis \( \{|g_{l}\rangle |g\in G\}\). The Hilbert space of the reversed edge ${l^{*}}$ is obtained via the isomorphism $\ket{g}_{l} \rightarrow \ket{g_{l^{*}}^{-1}}$. The gauge transformation operator at vertex \(v\) is given by
\begin{align}
    \mathcal{G}_{v}(g) = \prod_{l \ni v} U^{(v)}_{l}(g),
\end{align}
where the product runs over all links connected to $v$ and $U^{(v)}_{l}(g)$ acts according to the orientation of $l$ relative to $v$:
\begin{align}
    U^{(v)}_{l}(g)\ket{h_l} = \begin{cases}
        \ket{(g^{-1}h)_l}  &\text{tail($l$)=$v$}, \\
        \ket{(hg)_l} &\text{head($l$)=$v$}
    \end{cases}
\end{align}
Physical states must be gauge invariant, namely $\mathcal{G}_{v}(g)\ket{\psi} = \ket{\psi}$ for all $g\in G$, due to the absence of probe $G$-charges. 
The gauge invariance condition can be ``solved" by transforming to the Fourier or ``electric" basis of the gauge theory. By the Peter-Weyl theorem, $\mathcal{L}^{2}(G)$ can be decomposed as $\mathcal{L}^{2}(G) =\bigoplus_{j}\mathcal{R}_{j}\otimes \mathcal{R}_{j^*}$. Thus, the electric basis of $\mathcal{H}_{l}$ is given by $\ket{j_l,\mu_l,\nu_l}$, where $j_l$ indexes the irrep and $\mu_l$ and $\nu_l$ index states in $\mathcal{R}_{j_l}$ and $\mathcal{R}_{j_l^{*}}$ respectively. 

The two bases are related to one another as follows: 
\begin{align}
    \ket{j,\mu,\nu} = &\sqrt{\frac{d_{j}}{|G|}} \sum_{g\in G}\rho_{j}(g)_{\mu\nu}\ket{g},  \\
     \ket{g} = \sum_{j}&\sqrt{\frac{d_{j}}{|G|}} \sum_{\mu,\nu = 1}^{d_{j}}\rho_{j}(g)^{*}_{\mu\nu}\ket{j,\mu,\nu},
\end{align}
where $d_{j}$ is the dimension of irrep $j$. The inverse transformation follows from the identity $\frac{1}{|G|}\sum_{j}d_{j}\Tr{\rho_{j}(g)} = \delta_{g,e}$. Furthermore, under the action of $U^{(v)}_{l}(g)$, the electric basis states transform as,
\begin{align}
    U^{(v)}_{l}(g)\ket{j_l,\mu_l,\nu_l} = \begin{cases}
        \rho(g)_{\mu_l\mu_l'}\ket{j_l,\mu'_l,\nu_l}  &\text{tail($l$)=$v$}, \\
       \rho(g)^{\dagger}_{\nu'_l\nu_l}\ket{j_l,\mu_l,\nu'_l} &\text{head($l$)=$v$}.
    \end{cases}
\end{align}
One can thus view the $\mu_l$ label as being attached to the tail of link $l$ and the $\nu_l$ label as being attached to the head of link $l$. The orientation preserving isomorphism in the electric basis is given by $ \ket{j,\mu,\nu}\rightarrow  \ket{j^{*},\nu,\mu}$.

An arbitrary state $\ket{\Psi} \in \bigotimes_{l}\mathcal{H}_{l}$ can now be expanded as 
\begin{align}
    \ket{\Psi} = \sum_{\{j_{l},\mu_{l},\nu_{l} \}} \Psi^{\textbf{j}}_{\boldsymbol{\mu,\nu}}
    \bigotimes_{l}\ket{j_{l},\mu_{l},\nu_{l} }.
\end{align}
Gauss' law implies that $\Psi^{\textbf{j}}_{\boldsymbol{\mu,\nu}}$ must be an invariant tensor under the action of $\mathcal{G}_{v}(g)$ for all vertices $v$. In particular, if links $l_1$, $l_2$ and $l_3$ meet at a trivalent vertex $v$, with say $l_1$ and $l_2$ outgoing and $l_3$ incoming, the condition that $\Psi^{\textbf{j}}_{\boldsymbol{\mu,\nu}}$ form a singlet under $\mathcal{G}_{v}(g)$ implies that $\Psi^{\textbf{j}}_{\boldsymbol{\mu,\nu}}$ can be factorized in terms of the Clebsch Gordon coefficients as follows:
\begin{align}
    \Psi^{...j_{l_{1}}j_{l_{2}}j_{l_{3}}...}_{...\mu_{l_{1}}\mu_{l_{2}}\mu_{l_{3}}...\nu_{l_{1}}\nu_{l_{2}}\nu_{l_{3}}...} \propto C_{\mu_{l_{1}}\mu_{l_{2}}\nu_{l_{3}}}^{j_{l_{1}}j_{l_{2}}j_{l_{3}}} \Psi^{...j_{l_{1}}j_{l_{2}}j_{l_{3}}...}_{...\mu_{l_{3}}...\nu_{l_{1}}\nu_{l_{2}}...} 
\end{align}
Thus gauge invariance completely fixes the $\mu$ and $\nu$ dependence of the wavefunction.
\begin{align}
    \Psi^{\textbf{j}}_{\boldsymbol{\mu,\nu}} = \bigg(\prod_{l_1,l_2,l_3 \ni v}C_{\mu_{l_{1}}\mu_{l_{2}}\nu_{l_{3}}}^{j_{l_{1}}j_{l_{2}}j_{l_{3}}}\bigg) \Phi^{\textbf{j}}.
\end{align}
The remaining  data is encoded in the coefficients $\Phi^{\textbf{j}}$ which defines a spin-network wavefunction on the reduced Hilbert space spanned by basis states of the form $\{\ket{j_{l}} \}$. Formally, one can define an isometric map from gauge-invariant wavefunctions $\Psi^{\textbf{j}}_{\boldsymbol{\mu,\nu}} $ in the full gauge theory Hilbert space to spin-network states $\Phi^{\textbf{j}}$ living in the reduced Hilbert space. The only residual constraint on the spin-network wavefunction is that the irreps meeting at each vertex must admit fusion to the trivial irrep. The state $\ket{\Phi_{g}}$ defined in~\eqnref{eq:gauge_wavefunction} satisfies this condition and therefore defines a valid spin-network state. An associated gauge-invariant wavefunction may be reconstructed isometrically by contracting the spin-network wavefunction with the appropriate Clebsch–Gordan intertwiners at each vertex. In what follows, we will suppress this distinction and simply refer to $\ket{\Phi_{g}}$ as the gauge wavefunction.

Finally, the geometry of the original circuit defines boundary conditions for \(\ket{\Phi_{g}}\). Initial and final time slices correspond to \emph{rough boundaries}, which can be viewed as edge modes sourcing electric flux lines. The left and right spatial boundaries are \emph{smooth}; they do not source electric flux lines. In a deconfined phase, flux lines can spread through the bulk, while in a confining phase the amplitude for long electric strings decays as $e^{-T_{\text{str}} l}$, suppressing flux penetration into the circuit bulk (an ``electric Meissner'' effect).

\subsection{Bulk-Boundary Correspondence}

\begin{figure}
    \includegraphics[width=0.95\linewidth]{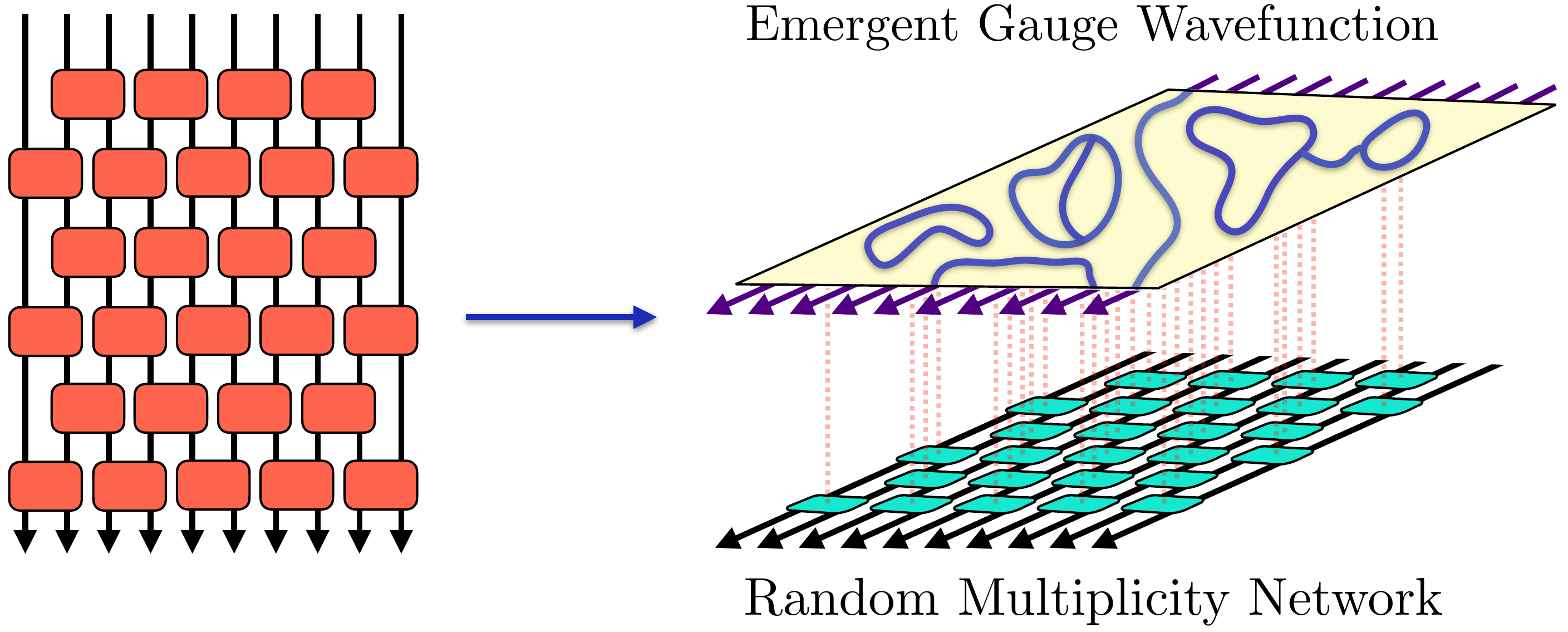}
    \caption{\label{fig:Bulk_Boundary_Correspondence}  {\bf Bulk-boundary correspondence.} The symmetric circuit, when viewed as a tensor network, decomposes into a fixed charged layer and an unconstrained multiplicity layer. The charged layer defines an emergent gauge wavefunction (``spin-network"). Contracting this bulk gauge wavefunction with random multiplicity tensors yields the boundary spin state.} 
\end{figure}
At each spacetime vertex \(v\), we define the multiplicity state
\begin{align}
    \ket{T_{v}} 
    = \sum_{\boldsymbol{j},\boldsymbol{a}} (T_{v})^{\boldsymbol{j}}_{\boldsymbol{a}}
      \bigotimes_{i=1}^{2k} \ket{j_{i},a_{i}},
\end{align}
where the tensor product runs over all the legs attached to \(v\). The full symmetric tensor network is given by the contraction of local multiplicity tensors with a fixed bulk spin-network state \(|\Phi_{g}\rangle \)~\cite{RTNs_2016, qi2022emergentbulkgaugefield}. This yields a natural two-layer decomposition of the symmetric tensor network, illustrated in~\figref{fig:Bulk_Boundary_Correspondence}. Note that multiplicity tensors reside on the vertices of the spacetime lattice, while the gauge state \(|\Phi_{g}\rangle \) is supported on the links. To contract them consistently, we introduce an isometry \(E\) that embeds the link Hilbert space into an \emph{extended Hilbert space} by assigning one copy of the link degree of freedom to each of its endpoint vertices. For an oriented link \(l: x \to y\), it is given by
\begin{align}\label{eq:Extended_Hilbert_Space_Map}
    E_{l}: \ket{j}_{l}
    \mapsto
    \frac{1}{\sqrt{n_{j}}} \sum_{a}
    \ket{j^{*},a}_{xy} \otimes \ket{j,a}_{yx},
\end{align}
where the charge at the beginning of the link is conjugated and the charge at the end is not. On links internal to the trivalent decomposition (which are not present in the original circuit), \(E_{l}\) is the identity. The full isometry is given by the product over all physical links,
\begin{align}
    E = \prod_{\text{links } l} E_{l}, \quad E^{\dagger} E = \mathbb{I}.
\end{align}
In gauge-theory language, \(E\) is known as the canonical embedding map into the extended Hilbert space~\cite{EE_GaugeTheory1,EE_GaugeTheory2}.


The boundary state of the tensor network at the initial and final timeslices is given by
\begin{align}\label{eq:Bulk_Boundary_Correspondence}
    \ket{\Phi_{\text{Bdy}}}
    = \Big(\hspace{-2pt} \otimes_{v} \hspace{-2pt} \langle T_{v} | \Big)  E \ket{\Phi_{g}}.
\end{align}
Appropriately contracting \( \ket{\Phi_{\text{Bdy}}}\) with the fixed initial state $|\psi_i \rangle$ yields the final state of the spin chain $|\psi_f \rangle$ in terms of the bulk gauge wavefunction and the random multiplicity tensors. This realizes a discrete bulk-boundary correspondence where the global symmetry of the boundary spin chain gives rise to gauge invariance in the bulk. The bulk direction in this case is identified with time.

We can generalize the bulk-boundary correspondence to circuits that include additional symmetry-preserving noise or measurement channels.  Consider a brickwork circuit built from \(G\)-symmetric gates, where after each layer of gates, a local quantum channel acts independently on each site,
\begin{align}
    \mathcal{E}^{t}(\cdot) = \prod_{k=1}^{L} \mathcal{E}^{t}_{k}(\cdot).
\end{align}
Each \(\mathcal{E}^{t}_{k}\) admits a Kraus representation of the form
\begin{align}
    \mathcal{E}(\cdot) = \sum_{\alpha} p_{\alpha} K_{\alpha}(\cdot) K_{\alpha}^{\dagger}.
\end{align}
We assume that the channel is strongly symmetric, namely all Kraus operators commute with the symmetry,
\begin{align}
    [K_{\alpha}, V(g)] = 0,
\end{align}
and acts as the identity on multiplicity indices. Examples include local charge-measurement channels.

In this setting, the bulk gauge state \(\ket{\Phi_{g}}\) is simply replaced by a mixed state
\begin{align}
    \rho_{g} = \prod_{\mu} \mathcal{E}_{\mu} \big( \ket{\Phi_{g}}\bra{\Phi_{g}} \big),
\end{align}
where \(\mathcal{E}_{\mu}\) now acts on the corresponding link degrees of freedom. Thus, local symmetric channels acting on the spin chain map to local channels on the bulk gauge wavefunction. The bulk-boundary correspondence now reads
\begin{align}
    \rho_{\text{Bdy}}
    = \Big( \bigotimes_{v} \bra{T_{v}} \Big)
      E \rho_{g} E^{\dagger}
      \Big( \bigotimes_{v} \ket{T_{v}} \Big).
\end{align}

\subsection{Unitarity and Deconfinement}\label{sec:Unitary_Deconfinement}

So far, we have not imposed any requirement on local gates besides symmetry. 
We now restrict our attention to $2\times2$ \emph{unitary} gates and examine what this implies for the structure of the gauge wavefunction. We first focus on the case where $G$ is abelian. In this setting, since all irreps are one-dimensional, the $\mu_j$ label can be dropped. Furthermore, all intermediate charge sectors are fixed and we may simply write
\begin{align}
    A^{j}_{\bm{a}} = B^{\bm{j}} T^{\bm{j}}_{\bm{a}}.
\end{align}
The $B^{\bm{j}}$ tensor is 1 if the charges obey the appropriate selection rule and is 0 otherwise. Any constant prefactor can be re-absorbed into $T^{\bm{j}}_{\bm{a}}$.

\subsubsection{$G=\mathbb{Z}_{N}$}
We first consider $\mathbb{Z}_{N}$-symmetric circuits with a local Hilbert space, $\mathcal{H} = \bigotimes_{j = 0}^{N-1} n R_{j}$. Here each one-dimensional irrep $R_{j}$ appears with multiplicity $n$. 
Separating the charge labels $\bm{j}$ into incoming and outgoing parts, $\bm{j}_{\text{in}}$ and $\bm{j}_{\text{out}}$, the $\mathbb{Z}_{N}$ charge conservation rule can be written as
\begin{align}
    \sum_{i} j_{\text{in},i} = \sum_{i} j_{\text{out},i} \pmod N.
\end{align}
Defining the total incoming and outgoing charges as $J_{\text{in}}  = \sum_{i} j_{\text{in},i}  $ and $J_{\text{out}}  = \sum_{i} j_{\text{out},i} $ respectively allows us to express the matrix elements of the gate in the form
\begin{align}
    A^{\bm{j}_{\text{out}}\bm{j}_{\text{in}}}_{\bm{a}_{\text{out}}\bm{a}_{\text{in}}} = \delta^{J_{\text{out}},J_{\text{in}}} T^{\bm{j}_{\text{out}}\bm{j}_{\text{in}}}_{\bm{a}_{\text{out}}\bm{a}_{\text{in}}}.
\end{align}
Thus, the gate is block-diagonal in total charge sectors, labeled by $J =J_{\text{out}} = J_{\text{in}}$. The total charge $J$ takes values $0, 1, \dots, N-1$, giving rise to $N$ independent charge sectors. Each block $T^{J}$ acts within a subspace of dimension $Nn^{2}$. For instance, for $\mathbb{Z}_{3}$, $U$ would look like
\[
  \setlength{\arraycolsep}{0pt}
  U = \begin{bmatrix}
    \fbox{$T^{0}_{3n^{2}\cross 3n^{2}}$} & 0 & 0  \\
    0 & \fbox{$T^{1}_{3n^{2}\cross 3n^{2}}$} & 0 \\
    0 & 0 & \fbox{$T^{2}_{3n^{2}\cross 3n^{2}}$} \\
  \end{bmatrix}.
\]
If each block $T^{J}$ is chosen to be unitary, the full gate $U$ is automatically unitary. Since we are interested in random circuit ensembles, we take the elements of each $T^{J}$ to be independently sampled from the Haar measure on $U(Nn^{2})$. Analytic treatment of such Haar-random ensembles is typically challenging. However, substantial simplifications occur in the large-multiplicity limit $n \gg 1$, where each block becomes large. In this regime, Haar averages coincide with Gaussian (Wick) averages owing to measure concentration on high-dimensional matrices~\cite{Entanglement_Dynamics_Membrane1}. In this limit, the matrix elements of each block are effectively independent Gaussian variables with mean zero and variance $1/Nn^{2}$.

Finally, note that since $B^{\bm{j}} = \delta^{J_{\text{out}}, J_{\text{in}}}$ (with the $\delta$-function understood modulo $N$), the corresponding gauge wavefunction $\ket{\Phi_\text{g}}$ takes the form of an equal-weight superposition over all $\mathbb{Z}_{N}$ string-net configurations compatible with the boundary conditions:
\begin{align}
    \ket{\Phi_\text{g}} = \sum_{\mathcal{C}}\ket{\mathcal{C}}.
\end{align}
This is precisely the deconfined fixed-point state of $\mathbb{Z}_{N}$ lattice gauge theory. The final state of the spin chain  $|\psi_f \rangle$ is obtained through the bulk-boundary mapping in \eqnref{eq:Bulk_Boundary_Correspondence}. 
Our objective is to evaluate non-linear functionals of the boundary density matrix ${|\psi_f\rangle\langle\psi_f|}$, averaged over the ensemble of multiplicity tensors. This requires computing averages over products of local tensors of the form $\overline{|T_{v}\rangle \langle T_{v}|^{\otimes k}}$. In the large bond-dimension limit, it simplifies to (see \appref{sec:Average_Multiplicity})
\begin{align}\label{eq:Average_Multiplicity}
    \overline{|T_{v}\rangle \langle T_{v}|^{\otimes k}} = \sum_{\sigma \in S_{k}}   M_v^{\otimes k}P_{\sigma} M_v^{\otimes k},
\end{align}
where $P_{\sigma}$ denotes the permutation operator acting on the $k$ replicated tensor copies, and $M_v = \frac{1}{\sqrt{Nn^{2}}} \mathbb{I}$. The net effect of this averaging procedure is to dress the gauge wavefunction by the operator $M = \prod_v M_v$, $\ket{\Phi_\text{g}}\rightarrow M \ket{\Phi_\text{g}}$. For $\mathbb{Z}_N$, the action of $M$ is purely multiplicative; it contributes only an overall normalization factor of $(Nn^{2})^{-V/2}$, where $V$ is the number of vertices in the lattice.

\subsubsection{$U(1)$ symmetric gates}
We next consider a $U(1)$-symmetric random circuit composed of two-site gates, with a local Hilbert space of the form $\mathcal{H} = \bigoplus_{j = 0}^{N-1}n\mathcal{R}_{j}$. Each local site carries an integer-valued charge $j \in {0, 1, \dots, N-1}$, with each one-dimensional charge sector appearing with multiplicity $n$. A key distinction between the continuous group $U(1)$ and the cyclic group $\mathbb{Z}_{N}$ is that $U(1)$, in principle, possesses an infinite tower of irreducible representations, whereas $\mathbb{Z}_{N}$ admits only finitely many. The corresponding charge-selection rule takes the simple form $J_{\text{out}} = J_{\text{in}}$. The total incoming/outgoing charge $J = J_{\text{out}} = J_{\text{in}}$ can therefore take values from $0$ to $2(N-1)$. The multiplicity of each total charge block is $D_{J}= n^{2} d_{J}$, where $d_{J}$ denotes the number of integer solutions to $j_{1} + j_{2} = J$ for $j_{1,2} \in {0, 1, \dots, N-1}$. As a concrete example, consider the qubit case ($N=2$), for which $\mathcal{H} = n\mathcal{R}_{-1}\oplus n\mathcal{R}_{1}$. Here the total charge $J$ can take values $-2$, $0$, or $2$, and the corresponding $U(1)$-symmetric two-site gate takes the block-diagonal form
\[
  \setlength{\arraycolsep}{0pt}
  U = \begin{bmatrix}
    \fbox{$T^{-2}_{n^{2}\cross n^{2}}$} & 0 & 0  \\
    0 & \fbox{$T^{0}_{2n^{2}\cross 2n^{2}}$} & 0 \\
    0 & 0 & \fbox{$T^{2}_{n^{2}\cross n^{2}}$} \\
  \end{bmatrix}.
\]
Unitarity of the gate requires that each charge block individually form a unitary matrix. Accordingly, we take each block to be independently drawn from the Haar measure on the unitary group $U(D_{J})$. As before, we work in the large-multiplicity limit $n \gg 1$, where each non-zero matrix element of the two-site gate can be effectively treated as an independent complex Gaussian random variable. Since the charge blocks have different dimensions, the Gaussian variances must be scaled appropriately for each block. In particular, the elements within the $J$-th charge block are assigned variances of $1/D_{J}$ to ensure proper normalization in the large-$n$ limit.

Again, since $B^{\bm{j}} = \delta^{J_{\text{out}}, J_{\text{in}}}$, the corresponding gauge wavefunction $\ket{\Phi_\text{g}}$ is an equal-weight superposition over all allowed electric string-net configurations consistent with the boundary conditions. The key difference here is that averaging over multiplicity tensors introduces additional factors which weigh the different string-net configurations differently. In the large bond-dimension limit, the average of the $k$th moment of the multiplicity state again takes the same form as in \eqnref{eq:Average_Multiplicity}
but $M_v$ now acts on the basis states $\ket{\bm{j},\bm{a}}$ in the following manner $M_v\ket{\bm{j},\bm{a}} = \frac{1}{\sqrt{D_{J_{\text{out}}}}}\ket{\bm{j},\bm{a}}$. Hence, the normalization factor of the gauge wavefunction at each vertex depends on the total charge entering or leaving it. 

\subsubsection{Finite non-abelian groups}
Finally, we consider a generic finite group $G$, which may be non-abelian. It possesses a finite set of irreps and the local Hilbert space is taken to be the direct sum of all irreps, each appearing with multiplicity $n$: $\mathcal{H} = \bigoplus_{j}n\mathcal{R}_{j}$. To analyze the constraints imposed by unitarity, we must first specify a normalization convention for the Clebsch-Gordan coefficients. We assume that these coefficients are normalized as follows,
\begin{align}\label{eq:CS_normalization}
    \sum_{j_{3},\mu_{3}} (C^{j_{1}j_{2}j_{3}}_{\mu_{1}'\mu'_{2}\mu_{3}})^{*} C^{j_{1}j_{2}j_{3}}_{\mu_{1}\mu_{2}\mu_{3}}  &= \delta_{\mu_{1}\mu_{1}'}\delta_{\mu_{2}\mu_{2}'}     \\
    \sum_{\mu_{1} \mu_{2}} (C^{j_{1}j_{2}j_{3}}_{\mu_{1}\mu_{2}\mu_{3}})^{*} C^{j_{1}j_{2}j'_{3}}_{\mu_{1}\mu_{2}\mu'_{3}}   &= \delta_{j_{3}j_{3}'}\delta_{\mu_{3}\mu_{3}'}.
\end{align}
For simplicity, we assume that $G$ is multiplicity-free, so that each irrep $j_{3}$ appears at most once in the fusion $j_{1}\otimes j_{2}$. Recall that the tensor components of a two-site $G$-symmetric gate decompose as
\begin{align}
    A^{J_{1}J_{2}J_{3}J_{4}} = \sum_{e} B^{j_{1}j_{2}j_{3}j_{4}; e }_{\mu_{1}\mu_{2}\mu_{3}\mu_{4}} T^{j_{1}j_{2}j_{3}j_{4}; e}_{a_{1}a_{2}a_{3}a_{4}}.
\end{align}
where the charged $B$ tensor is constructed from pairs of Clebsch-Gordan coefficients as follows,
\begin{align}
    B^{j_{1}j_{2}j_{3}j_{4}; e }_{\mu_{1}\mu_{2}\mu_{3}\mu_{4}} = \sum_{\mu_{e}} \big(C^{j_{1}j_{2}e}_{\mu_{1}\mu_{2}\mu_{e}}\big)^{*}C^{j_{3}j_{4}e}_{\mu_{3}\mu_{4}\mu_{e}}.
\end{align}
Again, the gate decomposes into independent blocks, one for each value of the total charge $e$.
\begin{align}\label{fig:QCMI_geometry}
        &\quad\vcenter{\hbox{\includegraphics[height=10ex]{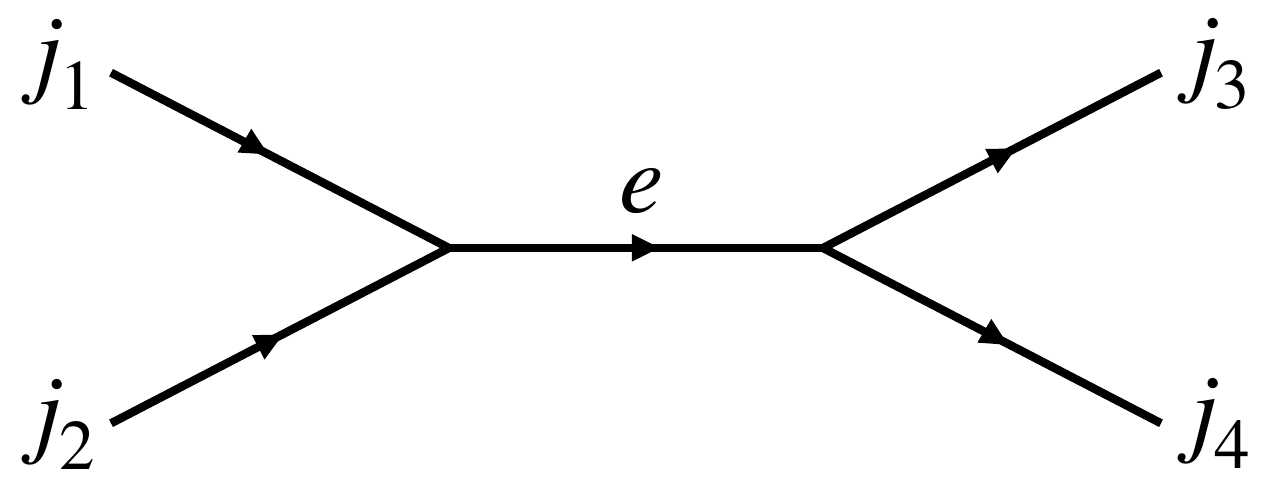}}} \nonumber
\end{align}
The dimension of this block is $D_{e} = n^{2} d_{e}$, where $d_{e}$ denotes the number of distinct fusion channels through which two incoming charges can combine to yield the intermediate charge $e$. We must again choose a suitable ensemble for the multiplicity tensor $T^{j_{1}j_{2}j_{3}j_{4}; e}_{a_{1}a_{2}a_{3}a_{4}}$ such that the resulting two-site gate is unitary. We show in \appref{app:Unitarity_nonabelian_gates} that unitarity of the full gate is achieved precisely when, for each intermediate charge $e$, the tensor $T^{j_{1}j_{2}j_{3}j_{4}; e}_{a_{1}a_{2}a_{3}a_{4}}$ is unitary in all charge and multiplicity indices belonging to that block. Therefore, unitarity again reduces to imposing blockwise unitarity within each total-charge sector. In the large-multiplicity limit, this is equivalent to sampling each tensor component in the charge $e$ block independently from a complex Gaussian distribution with mean zero and variance $1/D_{e}$. In this limit, the average of the $k$-th moment of the multiplicity state takes the same form as in \eqnref{eq:Average_Multiplicity}, except that the operator $M_v$ now acts on the trivalent vertex states $\ket{j_{1},a_{1}}\ket{j_{2},a_{2}}\ket{e}$ in the following manner $M_v\ket{j_{1},a_{1}}\ket{j_{2},a_{2}}\ket{e} = \frac{1}{\sqrt{D_{e}}}\ket{j_{1},a_{1}}\ket{j_{2},a_{2}}\ket{e}$. (Here we assume that the lattice has been decomposed into a trivalent network.) Once again, this contributes a charge-dependent normalization factor to the gauge wavefunction at each vertex. Importantly, however, the resulting gauge states still involve a coherent sum over all spin-network configurations consistent with the boundary conditions, and are therefore long-range entangled.

\subsection{Weak Measurements}

Finally, we incorporate measurements into the analysis. We focus on charge measurements whose Kraus operators $K_{s}$ commute with the group action, namely
\begin{align}
    [K_{s},V(g)] = 0.
\end{align}
This condition ensures that $K_{s}$ can distinguish only the irrep labels, not the internal state labels within each irrep. Such measurements exclusively modify the gauge wavefunction $\ket{\Phi_\text{g}}$ and leave the random multiplicity network unaltered. 
For example, in $SU(2)$ symmetric circuits, one may measure the Casimir operator, $L^{2} = L^{2}_{x} + L^{2}_{y} + L^{2}_{z}$, which determines the total-spin (irrep) label, but measuring any individual component $L_{i}$ would explicitly break the $SU(2)$ symmetry.

In general, measurements may be weak, with Kraus operators of the form
\begin{align}
    K_{s} = \sum_{j,\mu_{j},a_{j}} \sqrt{p_{j;s}} | j,\mu_{j},a_{j}\rangle \langle j,\mu_{j},a_{j}|.
\end{align}
The coefficients $p_{j;s} \geq 0$ quantify the strength of the measurement and they must sum to unity to ensure that $\sum_{s} K_{s}^{\dagger} K_{s} = \mathbb{I}$. Projective measurements correspond to the limit $p_{j;s} = \delta_{j,s}$. 
We consider weak measurements performed on each spin after every layer of unitary gates. For a given measurement history $\{s_{i}(t) \}$, where $i$ labels the spin and $t$ the timestep, the updated gauge state is given by
\begin{align}
    \prod_{\mu}\tilde{K}_{s_{\mu}} | \Phi_g \rangle   \quad \text{where } \tilde{K}_{s} :=  \sum_{j} \sqrt{p_{j;s}} |j\rangle \langle j|.
\end{align}
Here the product runs over all links of the spacetime lattice since measurements are applied between successive layers of unitaries which define the lattice vertices. Note that we've dropped the multiplicity labels in $K_{s}$ to define $\tilde{K}_s$.

For $G = \mathbb{Z}_{N}$, the Kraus operators are simply given as 
\begin{align}\label{eq:Zn_weak_measurement}
    \tilde{K}_{s} = \sum_{j=0}^{N-1} \sqrt{p_{j;s}}|j\rangle \langle j|.
\end{align}
When the charge labels $j$ are viewed as evenly spaced points on a circle, it is natural to impose $\mathbb{Z}_{N}$ translation invariance, $p_{j+k;s+k} = p_{j;s}$, which forces the parameters $p_{j;s}$ to only depend on the relative angular separation between $j$ and $s$, $p_{j;s} = p_{j-s}$. Furthermore, we will focus on \textit{clock type} weak measurements which are characterized by the additional charge-conjugation symmetry $p_{k}= p_{-k\mod N}$. This ensures that the measurement uncertainty is distributed symmetrically around the mean value.

\section{Dynamically generated Topological Codes}

In the previous section, we demonstrated how a gauge wavefunction $\ket{\Phi_{g}}$ naturally emerges in the description of random symmetric quantum circuits. 
We now demonstrate that distinct global charge sectors of the underlying spin chain map directly onto distinct topological sectors of the gauge theory. 
For $\mathbb{Z}_N$-symmetric circuits, subject to symmetry preserving noise channels, we show that the resulting bulk gauge theory realizes a noisy $\mathbb{Z}_N$ surface code. 
This allows us to identify a distinguished set of logical states of the spin chain that inherit the topological protection of the bulk surface code. We make this correspondence precise by demonstrating the equivalence of bulk and boundary coherent information.


\subsection{Global charge becomes topological}
Consider a $\mathbb{Z}_N$ symmetric spin chain. All group elements $g\in G$ can be written as powers of a single generator $a$ with $a^N = e$. 
To proceed, we define generalized $\mathbb{Z}_N$ Pauli operators $X$ and $Z$ which obey $X^N\,{=}\,Z^N\,{=}\,1$ and $XZ = \omega ZX$, where $\omega = e^{2\pi i/N}$.
Then, the symmetry action $V(a)$ can be written as 
\begin{align}
    V(a) = \prod_{k = 1}^{L} Z_{k} =:Z_{\text{tot}}.
\end{align}
The global Hilbert space decomposes into superselection sectors labeled by the total conserved charge, $\mathcal{H}^{\otimes L} =  \bigoplus_{Q=0}^{N-1} \mathcal{H}_{Q}$, where $Q$ denotes the global charge sector. Within each global charge sector, the symmetry operator $Z_{\text{tot}}$ acts as multiplication by a phase. 

\begin{figure}[!t]
    \includegraphics[width=0.95\linewidth]{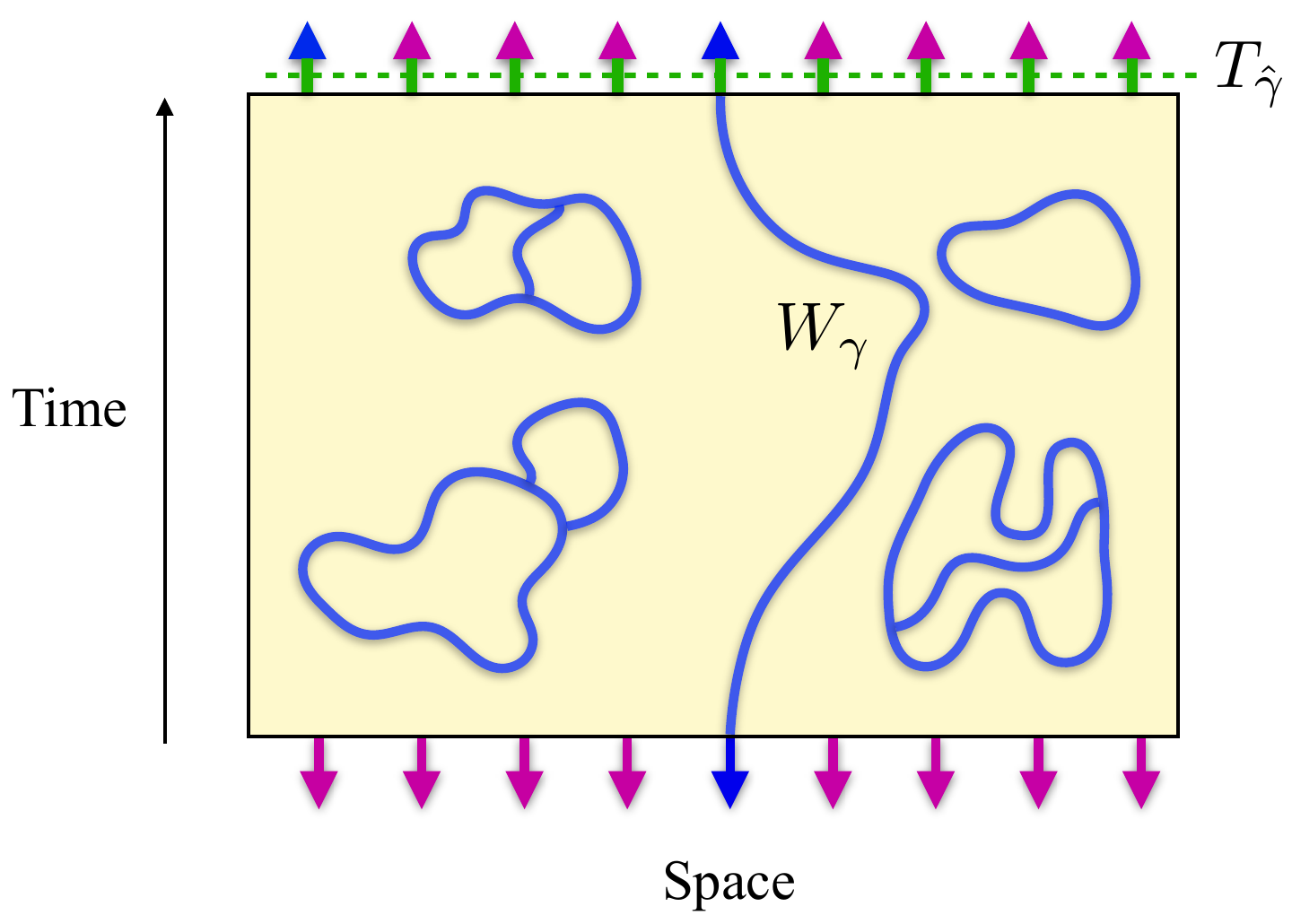}
    \caption{\label{fig:Loops}  {\bf Wilson and 't Hooft loops.} The ’t~Hooft operator, $T_{\hat{\gamma}}$, associated with a dual-lattice loop $\hat{\gamma}$ running along the rough boundary at the final timeslice
    generates a 1-form symmetry in the gauge theory and serves to distinguish different topological sectors. States in different topological sectors are related to one another through the action of a Wilson string, $W_{\gamma}$, which connects the rough boundaries at the initial and final timeslices.}
\end{figure}

Recall that the bulk-boundary correspondence (see~\eqnref{eq:Bulk_Boundary_Correspondence}) expresses the final state of the spin chain in terms of the gauge wavefunction and random multiplicity tensors. Employing the explicit form of the extended Hilbert space isometry $E$ (see~\eqnref{eq:Extended_Hilbert_Space_Map}), we find that the action of $Z_{\text{tot}}$ on the final spin-chain state can equivalently be represented as
\begin{align}\label{eq:Intertwining_identity}
    \prod_{k = 1}^{L} Z_{k}\bigg[\bigotimes_{v}(\bra{T_{v}} )E\ket{\Phi_\text{g}}\bigg] = \bigotimes_{v}\bra{T_{v}} ET_{\hat{\gamma}}\ket{\Phi_\text{g}} 
\end{align} 
where
\begin{align}
    T_{\hat{\gamma}} = \prod_{\mu = 1}^{L} Z_{\mu}
\end{align}
denotes the ’t~Hooft operator associated with a dual-lattice loop $\hat{\gamma}$ running along the rough boundary at the final timeslice (see \figref{fig:Loops}). When periodic boundary conditions are imposed on the spin chain, the spacetime lattice on which the gauge state is defined takes the form of a cylinder, and the ’t~Hooft operator $T_{\hat{\gamma}}$ corresponds to the familiar magnetic operator that winds around the non-contractible cycle of this cylinder.

Recall that in 2+1D gauge theory, a ’t~Hooft operator $T_{\hat{\mathcal{C}}}$ associated with a closed dual lattice loop $\hat{\mathcal{C}}$ measures the net electric flux crossing the loop. In pure gauge theory without charge defects, Gauss’s law enforces that the net electric flux through any {contractible} loop must vanish. However, it imposes no constraint on the flux through a {non-contractible} loop. States carrying different amounts of electric flux through homologically non-trivial loops belong to distinct topological superselection sectors of the gauge theory Hilbert space. 
These sectors are related by ``large gauge transformations" and cannot be connected by any local operator.

Since the ’t~Hooft operator corresponds to a global symmetry operator of the spin chain, it follows that the distinct topological sectors of the gauge theory correspond to different global charge sectors of the spin chain. This exemplifies a well-known phenomenon that when a 1-form symmetry operator in the bulk gauge theory is pushed to the boundary, it manifests as a 0-form global symmetry in the edge theory.

\subsection{Logical States}

For $\mathbb{Z}_{N}$ symmetric circuits, the structure of the bulk gauge wavefunction depends on two ingredients: the $\mathbb{Z}_N$ charge selection rule, and the initial spin state $|\psi_{i} \rangle$. If the initial state has support across multiple global charge sectors, the resulting gauge wavefunction involves a coherent superposition of distinct topological sectors. 

This observation allows us to encode a single $N$-dimensional qudit in the initial spin state using a basis with one state within each global charge sector. Under the bulk–boundary correspondence, this corresponds to topologically encoding a single qudit in the emergent gauge wavefunction. Since symmetry-preserving noise channels acting on the spin chain at each timestep map to local noise channels acting on the gauge wavefunction, the bulk theory dynamically realizes a noisy $\mathbb{Z}_{N}$ surface code. Using this insight, we identify a distinguished basis of initial states that is maximally robust against such noise, and define the logical space of the spin chain as the subspace spanned by these states. This allows us to interpret the $\mathbb{Z}_N$ symmetric circuit interspersed with $\mathbb{Z}_N$ symmetric noise channels, within the volume law phase, as a topological error correcting code: the logical information encoded in the initial state inherits the intrinsic topological protection of the bulk code. 


To define the logical subspace of the spin chain, we begin by considering product states with definite local charges:
\begin{align}
    \ket{\psi} = \ket{j_{1}} | j_2 \rangle \cdots \ket{j_{L}}.
\end{align}
Here we've dropped multiplicity labels as they play no role in the following discussion. Under symmetry-preserving noise, quantum information stored in superpositions of such states is rapidly destroyed. By contrast, consider preparing the initial spin-chain state as a uniform superposition over all simultaneous $Z$-eigenstates with fixed total charge $Q$, 
\begin{align}\label{eq:spin_logical_states}
    \ket{\psi(Q)} = \frac{1}{N^{(L-1)/2}} \hspace{-15pt} \sum_{\substack{j_1\cdots j_L \\Q \equiv \sum_{i}j_i \textrm{ mod $N$}}} \hspace{-15pt} \ket{j_{1}} \cdots \ket{j_{L}}.
\end{align}
The resulting gauge state $\ket{\Phi_{g}(Q)}$ is then a uniform superposition over all $\mathbb{Z}_N$ string-net configurations with initial and final time slices subject to the condition $T_{\hat{\gamma}}\ket{\Phi_g(Q)} = e^{2\pi iQ/N}\ket{\Phi_g(Q)}$. States with different $Q$s are only distinguished by the total $\mathbb{Z}_{N}$ flux that crosses any surface at intermediate circuit time. These are precisely the logical states of the $\mathbb{Z}_N$ surface code. Quantum information encoded in superpositions of these states is maximally protected against local symmetry preserving noise.  Different logical states are related to one another as follows,
\begin{align}
    \ket{\Phi_g(Q)} = W_\gamma^{Q}\ket{\Phi_g(0)},
\end{align}
where $W_\gamma$ is a Wilson loop (a product of $X$ operators on the direct lattice) along a non-contractible open curve $\gamma$ connecting the two rough boundaries (see \figref{fig:Loops}).

Accordingly, we define the logical subspace of the spin chain as,
\begin{align}\label{eq:logical_subspace}
   \mathcal{H}_{\text{logical}} =  \mathrm{span}\{\ket{\psi(Q)} \,|\, Q=0,1,...,N-1\}.
\end{align}
This logical subspace is protected against local noise upto a finite threshold since the corresponding bulk topological code, subject to the same noise channel, is expected to exhibit a finite noise threshold. More explicitly, suppose the symmetric circuit is interspersed with local symmetry-preserving quantum channels acting on the charge degrees of freedom, $\mathcal{E}^{t}(\cdot) = \prod_{k=1}^{L} \mathcal{E}^{t}_{k}(\cdot)$. Under the bulk–boundary correspondence, this induces a corresponding channel acting independently on each link of the gauge-theory wavefunction, $\prod_{\mu} \mathcal{E}_{\mu}\big(\cdot\big)$. 
Since the logical qudit of the surface code is stable against sufficiently weak local noise~\cite{Dennis_2002_Error_Threshold,Fan_Bao_Altman, Lee_Exact_Diagonalization, Lee_CSS_Codes, Vijay2025_appear}, the corresponding logical spin states likewise retain their encoded information in this regime.

We finally note that the distinction between product and scrambled states in their susceptibility to charge measurements
can be understood as the coherent counterpart of the distinction between mixed states with and without strong-to-weak spontaneous symmetry breaking (SWSSB)~\cite{Lee2023, StW_U(1), lessa2024strongtoweakspontaneoussymmetrybreaking, sala2024spontaneousstrongsymmetrybreaking, kim2024errorthresholdsykcodes, gu2024spontaneoussymmetrybreakingopen, Sanjay_hydrodynamics, Sanjay_SWSSB}. A product state with definite local charges is easily learned by charge measurements, whereas in a scrambled state, the same charge information is highly delocalized and therefore difficult to access. Put differently, these states exhibit sharply different responses to a (weakly symmetric) measurement channel~\cite{Lee2023}. In our setup, the circuit is initialized in a coherent SWSSB state, namely a pure state in which the charge is highly scrambled and thus enjoys enhanced protection against charge measurements~\footnote{We note that such a pure state, in fact, has a genuine SSB as well}. From this perspective, the charge purification transition can be understood as a restoration of strong symmetry.

\subsection{Coherent Information}
\begin{figure}
    \includegraphics[width=0.95\linewidth]{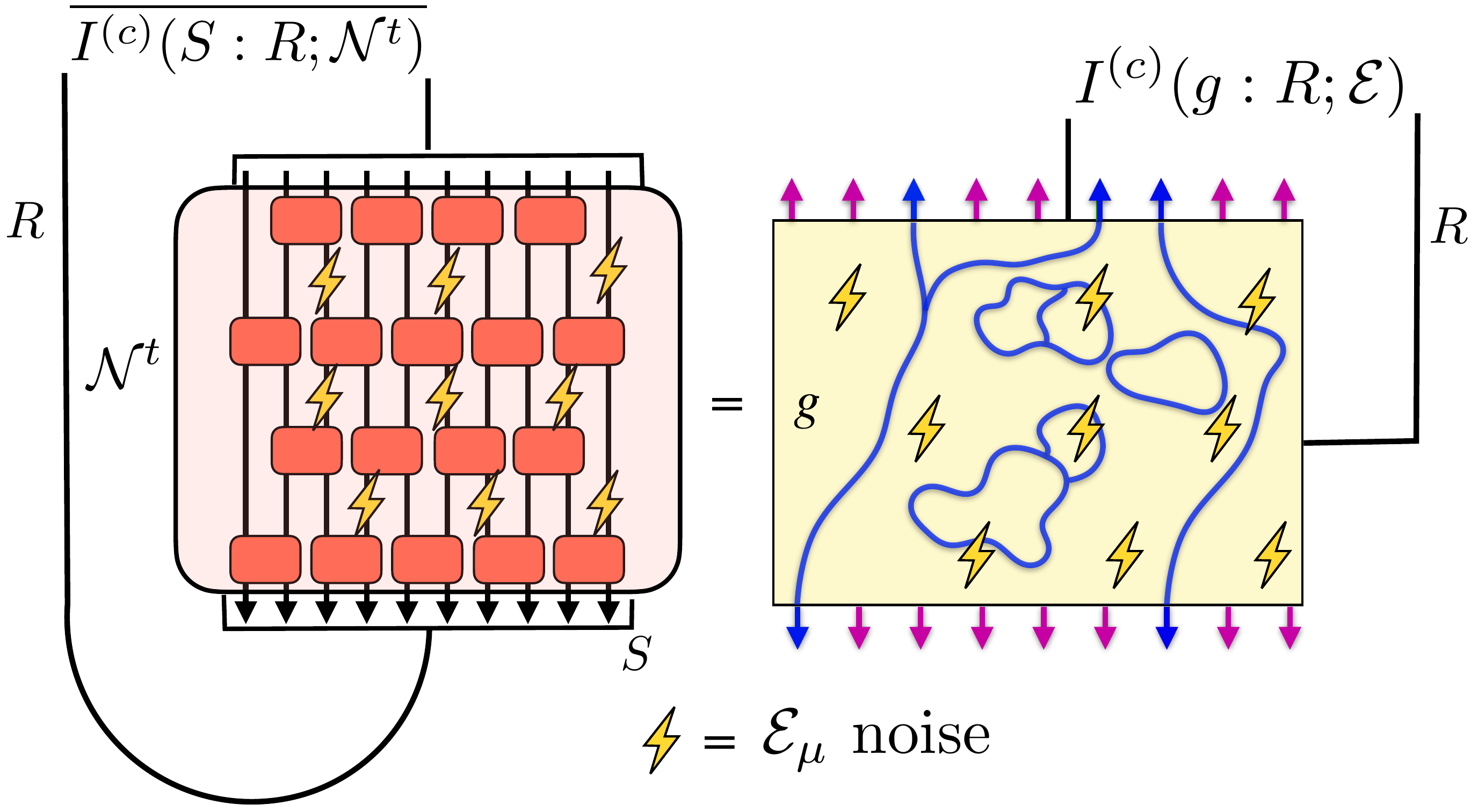}
    \caption{\label{fig:Coherent_Information}  {\bf Equality of Coherent Information:} Upon averaging over the multiplicity tensors in the large multiplicity limit, the coherent information of the spin chain evolving under the noisy symmetric circuit is precisely equal to the coherent information of the bulk surface code subjected to the same local noise.} 
\end{figure}

Let us examine the robustness of the quantum information encoded in the logical spin states $\{ \ket{\psi(Q)}\}$ in more detail. To this end, it is useful to imagine coupling the spin chain $S$ to an external reference $R$ whose Hilbert space dimension equals $N$, the number of distinct global charges of the spin chain. We then prepare the composite $SR$ system in the maximally entangled pure state
\begin{align}
    \ket{\Psi_{i}}_{SR} = \frac{1}{\sqrt{N}}\sum_{Q=0}^{N-1}\ket{\psi(Q)}_{S}\otimes \ket{Q}_{R}.
\end{align}
The reduced density matrix of $S$ is then just a classical mixture of all logical states
\begin{align}\label{eq:Initial_density_matrix}
    \rho_{S} =\frac{1}{N} \sum_{Q}|{\psi(Q)}\rangle_{S} \langle \psi(Q)|_{S}.
\end{align}
The spin chain is then evolved using local unitary gates arranged in a bricklayer architecture, interleaved with local noise channels of the form $\mathcal{E}^{t}(\cdot) = \prod_{k=1}^{L} \mathcal{E}^{t}_{k}(\cdot)$. Let us denote the combined quantum channel after $t$ time steps by $\mathcal{N}^{t}$ (see~\figref{fig:Coherent_Information}). The amount of distillable quantum information that can be extracted following the action of $\mathcal{N}^{t}$ is quantified by the coherent information~\cite{Lloyd_Coherent_Information,Shor_Coherent_Information,Devetak_Coherent_Information,Nielsen_Chuang}, which is defined as
\begin{align}
    I^{(c)}(S:R;\mathcal{N}^{t}) = S(\mathcal{N}^{t}(\rho_{S})) -  S(\mathcal{N}^{t}(\rho_{SR})),
\end{align}
where $S(\cdot)$ denotes the von Neumann entropy. Since the coherent information obeys the data processing inequality, $I^{(c)}(S:R;\mathcal{N}^{1}\circ \mathcal{N}^{2}) \leq I^{(c)}(S:R;\mathcal{N}^{2})$, the amount of information that is lost as a consequence of noise can never be recovered by post-processing. Furthermore, as a consequence of the sub-additivity of entropy, it is easy to see that
\begin{align}
    -\log N \leq I^{(c)}(S:R;\mathcal{N}^{t}) \leq \log N.
\end{align}
A positive coherent information heralds distillable entanglement whereas a negative coherent information points to consumption of entanglement by the noisy channel. 

Using the bulk boundary correspondence, we can write the final state of the spin chain as
\begin{align}
   \mathcal{N}^{t}(\rho_{S}) &= \bigg(\bigotimes_{v}\bra{T_{v}} \bigg)E \rho_{g} E^{\dagger} \bigg(\bigotimes_{v}(\ket{T_{v}}\bigg), \nonumber\\
    \mathcal{N}^{t}(\rho_{SR}) &= \bigg(\bigotimes_{v}\bra{T_{v}} \bigg)E \rho_{gR} E^{\dagger} \bigg(\bigotimes_{v}(\ket{T_{v}}\bigg),
\end{align}
where $\rho_{gR} = \prod_{\mu}\mathcal{E}_{\mu}(|\Phi_{{g}R}\rangle \langle \Phi_{{g}R}|)$, the state $\ket{\Phi_{{g}R}}$ is given by
\begin{align}
    \ket{\Phi_{{g}R}} = \frac{1}{\sqrt{N}}\sum_{Q=0}^{N-1}\ket{\Phi_{g}(Q)}\otimes \ket{Q}_{R},
\end{align}
and $\rho_{g} = \Tr_{R}\rho_{gR}$.
Note that $\ket{\Phi_{{g}R}}$ is just a maximally entangled state between the logical subspace of the dynamically generated surface code and the reference system.

For now, let us consider the case where the channels $\mathcal{E}(\cdot)$ represent decoherence channels rather than measurement channels, so we do not have access to the classical information carried by the measurement register. We will investigate measurement channels in the next section. Our goal is to compute the coherent information averaged over realizations of the random multiplicity tensors. In the large-multiplicity limit, namely deep within the volume-law phase, we show in \appref{sec:entropy_equality} that
\begin{align}
    \overline{S(\mathcal{N}^{t}(\rho_{S}))} &= S(\mathcal{E}(\rho_{g})), \nonumber \\
    \text{and}\quad\overline{S(\mathcal{N}^{t}(\rho_{SR}))} &= S(\mathcal{E}(\rho_{gR})). 
\end{align}
Namely, the averaged von Neumann entropy of the spin state evolved by the quantum channel $\mathcal{N}^{t}$ equals the bulk von Neumann entropy of the corresponding surface-code state subject to the linkwise decoherence channel $\mathcal{E} = \prod_{\mu}\mathcal{E}_{\mu}$. Consequently, the averaged coherent information of the $\mathbb{Z}_{N}$ symmetric spin chain is precisely equal to the coherent information of the bulk $\mathbb{Z}_{N}$ surface code:
\begin{align}
    \overline{ I^{(c)}(S:R;\mathcal{N}^{t})} &=  I^{(c)}(g:R;\mathcal{E}). 
\end{align}
For the $\mathbb{Z}_N$ surface code, it is well established that the coherent information remains approximately constant at $\log N$ in the regime of weak decoherence, reflecting perfect recoverability of the encoded logical information. This robustness is directly inherited by the spin chain and the logical qudit encoded in the spin states $\{ \ket{\psi(Q)}\}$ remains protected against local errors induced by decoherence.

\section{$\mathbb{Z}_{N}$ Charge Sharpening Transition}
Charge sharpening is a measurement-induced phase transition in which an initial spin state, prepared as a superposition of multiple global charge sectors, collapses onto a single, definite charge sector once the (rate) strength of (projective) weak measurements becomes sufficiently large. A natural diagnostic for detecting this transition is the average variance of the global charge or symmetry operator. For $G = \mathbb{Z}_{N}$, each element of the group is generated by a single element $a$ with $a^{N} = e$. Thus, it suffices to consider the variance of $V(a) = \prod_{k = 1}^{L}Z_{k} =:Z_{\text{tot}}$ where $Z_{k}$ is the generalized Pauli operator at site $k$. Since $Z_{\text{tot}}$ is unitary, its variance at time $t$, conditioned on measurement outcomes $\bm{s} = \{ s_{\mu}\}$, is simply given by
\begin{align}
    \textrm{Var}[Z_{\text{tot}}]^{\bf{s}}_t :&= \langle |Z_{\text{tot}} -  \langle Z_{\text{tot}} \rangle_{\mathbf{s},t}|^{2}\rangle_{\mathbf{s},t} \nonumber \\
    &= 1 - |\langle Z_{\text{tot}}\rangle_{\mathbf{s},t} |^{2},
\end{align}
where $\langle \cdot \rangle_{\mathbf{s},t}$ denotes the normalized expectation value with respect to the spin chain state $\ket{\psi^{\mathbf{s}}(t)}$ at time $t$ for a specific trajectory of measurement outcomes $\bm{s}$. If the initial state $\ket{\psi_i}$ is chosen to maximize the charge variance $\textrm{Var}[Z_{\text{tot}}]_0 =1$, then charge sharpening occurs when $\textrm{Var}[Z_{\text{tot}}]^{\bf{s}}_t \approx 0$, indicating that the spin chain has collapsed to a definite global charge sector. Conversely, if $\textrm{Var}[Z_{\text{tot}}]^{\bf{s}}_t = \cO(1)$, the final state retains multiple charge sectors and the global charge remains fuzzy. 

This setup involves two layers of randomness: (1) the local multiplicity tensors of the circuit are independently drawn from a random ensemble, and (2) the measurement outcomes $\bm{s}$ are random in accordance with the Born rule. Since we are interested in typical behaviour of the system, we need to average over both sources of randomness but the order of averaging is crucial since the Born probabilities themselves depend on the sampled multiplicity tensors.

We first average $\textrm{Var}[Z_{\text{tot}}]^{\bf{s}}_t$ over the post measurement ensemble $\{ \ket{\psi^{\bf{s}}(t)}\}$. Each such trajectory occurs with probability $p^{\bf{s}}(t) = \braket{\psi^{\bf{s}}(t)}{\psi^{\bf{s}}(t)}$. 
In the case of projective measurements, we must average over both the measurement locations and the measurement outcomes, whereas in the case of weak measurements, we only average over the measurement outcomes. We denote the average over measurement outcomes $\bm{s}$ using double brackets:
\begin{align}
    \lAngle \textrm{Var}[Z_{\text{tot}}] \rAngle_{t} &= \sum_{\bf{s}}p^{\bf{s}}(t) \textrm{Var}[Z_{\text{tot}}] ^{\textbf{s}}_t \nonumber \\
    &= 1 -  \sum_{\bf{s}}p^{\bf{s}}(t) |\langle Z_{\text{tot}}\rangle_{\mathbf{s},t} |^{2}.
\end{align}
We then perform a second average over the ensemble of random multiplicity tensors, denoted by $\overline{\cdots}$. The resulting quantity is:
\begin{align}\label{eq:Z_Variance}
   \overline{\lAngle \textrm{Var}[Z_{\text{tot}}] \rAngle}_{t} = 1 - \overline{\lAngle|\langle Z_{\text{tot}}\rangle_{\mathbf{s},t} |^{2}\rAngle}.
\end{align}
Since this quantity involves a measurement average of a non-linear function of the state, it is capable of detecting the charge sharpening transition. In particular, in the charge fuzzy phase, we expect that $\overline{\lAngle \textrm{Var}[Z_{\text{tot}}] \rAngle}_{t} = \mathcal{O}(1)$ upto timescales extensive in system size whereas in the charge-sharp phase, we expect that $\overline{\lAngle \textrm{Var}[Z_{\text{tot}}] \rAngle}_{t}  \approx 0$ already at $\mathcal{O}(1)$ time. Evidently, an equivalent diagnostic for the sharpening transition is the quantity 
\begin{align}\label{eq:Order_parameter}
    \overline{\lAngle|\langle Z_{\text{tot}}\rangle_{\mathbf{s},t} |^{2}\rAngle}
     = \sum_{\bf{s}}\overline{\Bigg[\frac{|\mel{\psi^{\mathbf{s}}(t)}{Z_{\text{tot}}}{\psi^{\mathbf{s}}(t)}|^{2}}{\braket{\psi^{\mathbf{s}}(t)}{\psi^{\mathbf{s}}(t)}}\Bigg]},
\end{align}
which displays precisely the opposite behaviour, namely it remains approximately $0$ upto timescales extensive in system size in the charge fuzzy phase whereas in the charge sharp phase it rapidly saturates to 1 on a timescale $t\sim \mathcal{O}(1)$. In the following, we demonstrate how this order parameter is evaluated within the gauge theory and explain the nature of the associated bulk transition.

\subsection{Gauge Theoretical Description}
Using the bulk-boundary correspondence, the final state of the spin chain may be expressed as $\ket{\psi^{\bf{s}}(t)} = \bigotimes_{v}(\bra{T_{v}} )E\ket{\Phi^{\bf{s}}_\text{g}}$ where $\ket{\Phi^{\bf{s}}_\text{g}} = \tilde{K}_{\bf{s}}\ket{\Phi_\text{g}}$ is the (unnormalized) gauge wavefunction conditioned on measurement outcomes $\bf{s}$. Using the intertwining identity in Eq.~\eqref{eq:Intertwining_identity}, the action of the symmetry operator $Z_{\text{tot}}$ on the spin chain can be replaced by that of a non-contractible ’t~Hooft loop $T_{\hat{\gamma}}$ which threads the rough boundary of the bulk gauge theory at the final timeslice. In \appref{sec:Sharpening_Diagnostic}, we show that averaging over the multiplicity tensors in the large-multiplicity limit yields the simple equation
\begin{align}\label{eq:averaged_variance}
    \overline{\lAngle |\langle Z_{\text{tot}}\rangle_{\mathbf{s},t} |^{2}\rAngle}   &=  \sum_{\bf{s}}p({\bf{s}})\Bigg|\frac{\bra{\Phi^{\boldsymbol{s}}_{\text{g}}} {T_{\hat{\gamma}}}\ket{\Phi^{\boldsymbol{s}}_{\text{g}}}}{\braket{\Phi^{\boldsymbol{s}}_{\text{g}}}{\Phi^{\boldsymbol{s}}_{\text{g}}}}\Bigg|^{2}  \nonumber \\
    &= \lAngle |T_{\hat{\gamma}}|^{2} \rAngle,
\end{align}
where $p({\bf{s}}) = \braket{\Phi^{\boldsymbol{s}}_{\text{g}}}{\Phi^{\boldsymbol{s}}_{\text{g}}}$. Thus, the diagnostic for sharpening is precisely the measurement average of the squared expectation value of a non-contractible ’t~Hooft operator. We choose the initial state to be the uniform superposition of logical spin states defined in~\eqref{eq:spin_logical_states}
\begin{align}
    \ket{\psi_i} = \frac{1}{\sqrt{N}}\sum_{Q = 0}^{N-1}\ket{\psi(Q)}.
\end{align}
This is an eigenstate of $X_{\text{tot}}$ with eigenvalue +1, satisfying $\textrm{Var}[Z_{\text{tot}}]_0 =1$. Since it is a logical state, it is robust to errors induced by measurements. The corresponding gauge state is given by 
\begin{align}
    \ket{\Phi^{\boldsymbol{s}}_\text{g}} = \sum_{Q = 0}^{N-1}\ket{\Phi^{\boldsymbol{s}}_\text{g}(Q)},
\end{align}
where $\ket{\Phi^{\boldsymbol{s}}_\text{g}(Q)} = \tilde{K}_{\boldsymbol{s}}\ket{\Phi_\text{g}(Q)}$ and we've dropped the normalization for convenience. The measurement Kraus operators commute with $T_{\hat{\gamma}}$, and since $T_{\hat{\gamma}}\ket{\Phi^{\boldsymbol{s}}_\text{g}(Q)} = \omega^{Q}\ket{\Phi^{\boldsymbol{s}}_\text{g}(Q)}$ with $\omega = e^{2\pi i/N}$, it follows that 
\begin{align}\label{eq:Order_Parameter}
    \lAngle  |T_{\hat{\gamma}}|^{2} \rAngle = \sum_{\bf{s}}\bigg(\sum_{Q' = 0}^{N-1}\mathcal{Z}[\textbf{s},Q'] \bigg)\Bigg|\frac{\sum_{Q = 0}^{N-1}\omega^{Q}\mathcal{Z}[\textbf{s},Q]}{\sum_{Q = 0}^{N-1}\mathcal{Z}[\textbf{s},Q] } \Bigg|^{2},
\end{align}
where we've defined $\mathcal{Z}[\textbf{s},Q] =\braket{\Phi^{\boldsymbol{s}}_{\text{g}}(Q)}{\Phi^{\boldsymbol{s}}_{\text{g}}(Q)}$.
These quantities correspond to the Born probabilities for observing measurement outcomes $\bf{s}$ in the logical state $\ket{\Phi_{g}(Q)}$. Below, we show that these probabilities can be naturally reinterpreted as partition functions of a disordered statistical-mechanics model.

\subsection{Local Charge Sharpening}
We can also consider the fluctuation of charge contained in a local region of the spin chain. For instance, consider a subregion $A$ of the spin chain of length $l_{A}$. The associated $\mathbb{Z}_{N}$ symmetry operator is $Z_{A} =  \prod_{k \in A}Z_{k}$ and its variance is given by
\begin{align}
    \mathrm{Var}[Z_A]^{\mathbf{s}} _t= 1 - |\langle Z_A\rangle_{\mathbf{s},t}|^2.
\end{align}
Once again, averaging the squared expectation $|\langle Z_A\rangle_{\mathbf{s},t}|^2$, first over all measurement outcomes, and then over the ensemble of random multiplicity tensors yields 
\begin{align}
    \overline{\lAngle |\langle Z_{\text{tot}}\rangle_{\mathbf{s},t} |^{2}\rAngle}   &=  \sum_{\bf{s}}p({\bf{s}})\Bigg|\frac{\bra{\Phi^{\boldsymbol{s}}_{\text{g}}} {T_{\hat{\gamma}_{A}}}\ket{\Phi^{\boldsymbol{s}}_{\text{g}}}}{\braket{\Phi^{\boldsymbol{s}}_{\text{g}}}{\Phi^{\boldsymbol{s}}_{\text{g}}}}\Bigg|^{2}  \nonumber \\
    &= \lAngle |T_{\hat{\gamma}_{A}}|^{2} \rAngle,
\end{align}
where $T_{\hat{\gamma}_{A}}$ is an open 't Hooft string inserted at the final timeslice which threads across subregion $A$. This can be interpreted as creating a pair of magnetic fluxes on the ends of subregion $A$. Thus, the expectation value $\langle T_{\gamma_{A}} \rangle$ is expected to behave as a two point correlator.

\subsection{Random Bond Clock Models}

In this section, we demonstrate how the probabilities $\braket{\Phi^{\boldsymbol{s}}_{\text{g}}(Q)}{\Phi^{\boldsymbol{s}}_{\text{g}}(Q)}$ can be re-expressed as partition functions for disordered clock models. We take the weak measurement Kraus operators to be of the form
\begin{align}
    \tilde{K}_{s} &= {\sum_{j = 0}^{N-1}\sqrt{p_{j;s}}|j\rangle \langle j|},
\end{align}
where the states $|j\rangle$ are again Pauli $Z$ eigenstates. The logical surface code state $\ket{\Phi_{g}(Q)}$ is a uniform superposition of all allowed string-net configurations satisfying $T_{\hat{\gamma}}\ket{\Phi^{\boldsymbol{s}}_\text{g}(Q)} = \omega^{Q}\ket{\Phi^{\boldsymbol{s}}_\text{g}(Q)}$. A convenient way to parametrize these states is in terms of a $\mathbb{Z}_{n}$-valued flow variable $\bf{k}$ defined on the links of the lattice, representing the eigenvalue of the Pauli $Z$ operator associated with each link. In this notation, we can write
\begin{align}
    \ket{\Phi_{g}(Q)} = \sum_{\substack{\bf{k}| \nabla\cdot \bf{k} = 0 \\ \hat{\gamma}(\bf{k}) = Q}} \ket{\bf{k}},
\end{align}
where the lattice divergence condition $\nabla\cdot \bf{k} = 0 \mod N$ enforces Gauss' law on each vertex and $\hat{\gamma}(\bf{k}) = Q$ fixes the $\mathbb{Z}_{n}$ flux threading the non-contractible cycle $\hat{\gamma}$ to be $Q$, namely $\sum_{\mu \perp \hat{\gamma}}k_{\mu} = Q\mod N$. Acting on this state with the weak measurement operators then yields
\begin{align}
    \prod_{\mu}\tilde{K}_{s_{\mu}}\ket{\Phi_{g}(Q)} &= \sum_{\substack{\textbf{k}| \nabla\cdot \textbf{k} = 0 \\ \hat{\gamma}(\textbf{k}) = Q}}  \prod_{\mu} \sqrt{p_{k_{\mu};s_{\mu}}}\ket{\bf{k}}.
\end{align}
It then follows that 
\begin{align}\label{eq:partition_function_current_loop}
     \mathcal{Z}[\textbf{s};Q] = \braket{\Phi^{\boldsymbol{s}}_{\text{g}}(Q)}{\Phi^{\boldsymbol{s}}_{\text{g}}(Q)} =\sum_{\substack{\textbf{k}| \nabla\cdot \textbf{k} = 0 \\ \hat{\gamma}(\textbf{k}) = Q}}  \prod_{\mu} p_{k_{\mu};s_{\mu}}.
\end{align}
For \textit{clock type} weak measurements characterized by the properties, $p_{j;s} \equiv p_{j-s\mod N}$ and $p_{k}= p_{-k\mod N}$, $\mathcal{Z}[\textbf{s};Q]$ resembles the current-loop representation of a $\mathbb{Z}_{N}$ clock model with quenched bond disorder $\boldsymbol{s} =\{ s_{\mu}\}$, restricted to a fixed homology sector labeled by $Q$. In fact, the dual \textit{height model} precisely assumes the form of a random phase clock model (``discrete gauge glass"). To see this, we first write $p_{k} = \frac{e^{\alpha_{k}}}{\mathcal{N}}$, where $\mathcal{N} = \sum_{k}e^{\alpha_{k}}$, and define the discrete $\mathbb{Z}_{N}$ Fourier transform of $\alpha_{k}$ as follows 
\begin{align}
    \alpha_{k} &= \sum_{m = 0}^{[N/2]}\beta_{m} \cos{\frac{2\pi km}{N}}.
\end{align}
Only cosine modes appear due to the assumed parity symmetry $p_{k}= p_{-k\mod N}$. It then follows that $p_{j;s} = \frac{1}{\mathcal{N}}\exp{\sum_{m = 0}^{[N/2]}\beta_{m} \cos{\frac{2\pi (j-s)m}{N}}}$. Next we solve the divergence free condition by introducing $\mathbb{Z}_{N}$ ``height variables" $\theta_{p}$ on each plaquette of the lattice and defining 
\begin{align}\label{eq:Dual_spin_definition}
    k_{\langle p_{1}\rightarrow p_{2}\rangle} &=  \theta_{p_{1}} - \theta_{p_{2}} \mod N.
\end{align}
Here $\langle p_{1}\rightarrow p_{2}\rangle$ denotes the unique oriented dual lattice link that cuts the direct lattice link $\mu$ from left to right. Such flows satisfy both $\nabla\cdot \bf{k} = 0 $ and $\hat{\gamma}(\bf{k}) = 0$. To generate a non-trivial $\mathbb{Z}_{n}$ flux through $\hat{\gamma}$, we may choose an arbitrary direct-lattice loop $\gamma$ connecting the initial and final timeslice boundaries and add a flow $Q$ along $\gamma$. This is equivalent to modifying the quenched background configuration by subtracting $Q$ from every link along $\gamma$
\begin{align}\label{eq:twisted_bonds}
    s^{Q}_{\mu} = \begin{cases}
        s^{0}_{\mu} - Q \quad \mu \in \gamma  \\
        s^{0}_{\mu}  \quad \mu \notin \gamma.
    \end{cases}
\end{align}
This then allows us to write (upto normalization)
\begin{align}\label{eq:Hamiltonian}
    \mathcal{Z}[\textbf{s}^{Q}] &=  \sum_{\{\theta_{p}\}}\exp{-H[\{\theta_{p}\};\textbf{s}^{Q}]},  \\
    H[\{\theta_{p}\};\textbf{s}^{Q}] &= -\sum_{\langle p_{1}, p_{2}\rangle} \sum_{m = 0}^{[N/2]}\beta_{m} \nonumber \\
    &\times\cos{\frac{2\pi m \big(\theta_{p_{1}} - \theta_{p_{2}}-s^{Q}_{\langle p_{1}, p_{2}\rangle}\big)}{N}}.
\end{align}
\begin{figure*}
    \includegraphics[width=0.95\textwidth]{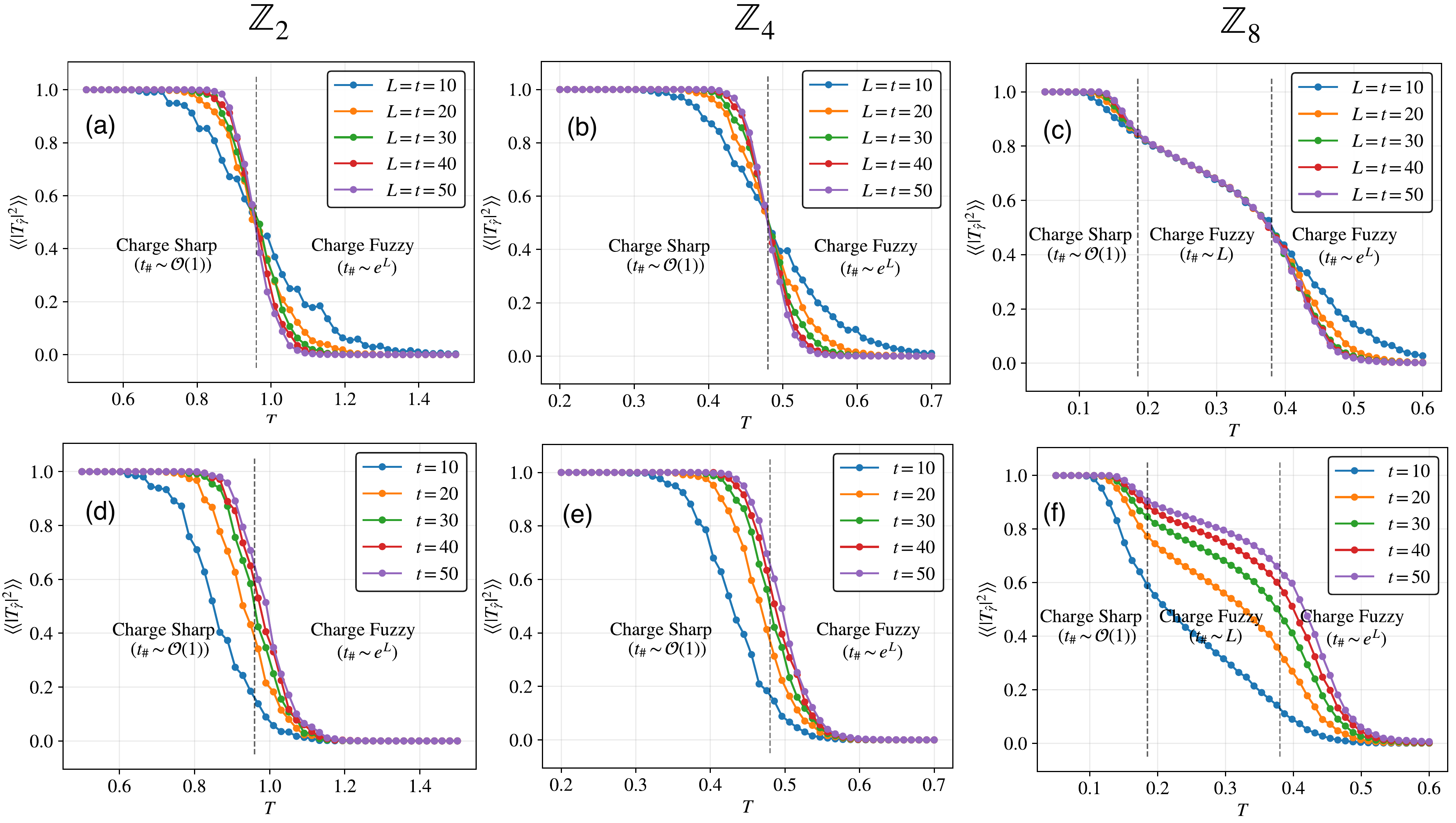}
    \caption{\label{fig:Numerics_Var}  {\bf Numerical results for $\langle |T_{\hat{\gamma}}|^{2} \rangle$:} We study charge sharpening dynamics in $\mathbb{Z}_{2}$ (left), $\mathbb{Z}_{4}$ (middle), and $\mathbb{Z}_{8}$ (right) symmetric circuits subject to the weak measurement model of Eq.~\eqref{eq:Measurement_model}.    The measurement strength is controlled by a single effective parameter $T$, which we interpret as a temperature. We evaluate $\langle |T_{\hat{\gamma}}|^{2} \rangle$ using the worm algorithm; this quantity directly diagnoses the sharpening transition. When  $\lAngle |T_{\hat{\gamma}}|^{2} \rAngle \approx 0$, the global charge remains fuzzy whereas when $\lAngle |T_{\hat{\gamma}}|^{2} \rAngle \approx 1$, the charge sharpens. Throughout $L$ denotes the size of the spin chain and $t$ denotes time. Figures (a), (b) and (c) plot $\lAngle |T_{\hat{\gamma}}|^{2} \rAngle$ for $L=t$. Note that for large $T$, when measurements are weak, the global charge remains fuzzy whereas for small $T$, when measurements are strong, global charge sharpens rapidly. Notably, in the $\mathbb{Z}_{8}$ case we find a clear regime where the curves collapse for all $L = t$, a hallmark of the QLRO (Coulomb) phase. Figures (d), (e) and (f) plot $\lAngle |T_{\hat{\gamma}}|^{2} \rAngle$ for fixed system size $L=30$ and varying circuit time $t$.}  
\end{figure*}
This is precisely the partition function of a $\mathbb{Z}_{N}$ clock model with quenched phase disorder $\{ s_{\mu}\}$ determined by the measurement outcomes and couplings $\beta_{m}$ determined by the measurement channel. Note that the net effect of non-trivial flux $Q$ is to impose a $(-Q)$-twisted boundary condition along the path $\gamma$ in the clock model. This introduces a form of topological frustration that energetically biases the system toward one topological sector over the rest.

Finally, note that the expectation value of an open 't Hooft string $T_{\hat{\gamma}_{A}}$ along subregion $A$ can be written as
\begin{align}
    \bra{\Phi^{\boldsymbol{s}}_{\text{g}}} {T_{\hat{\gamma}_{A}}}\ket{\Phi^{\boldsymbol{s}}_{\text{g}}} = \sum_{\{\theta_{p}\}} \omega^{\theta_{p_{1}} - \theta_{p_{2}}} e^{-H[\{\theta_{p}\};\textbf{s}^{Q}]},
\end{align}
where ${p_{1}}$ and $p_{2}$ are the plaquettes at the endpoints of the dual lattice curve $\hat{\gamma}_{A}$. This is a (disconnected) spin–spin correlation function in the random phase clock model for a fixed realization of the disorder $\bf{s}$. It follows that
\begin{align}\label{eq:Local_Sharpening_Order_parameter}
    \lAngle |T_{\hat{\gamma}_{A}}|^{2} \rAngle = [\langle \omega^{\theta_{p_{1}} - \theta_{p_{2}}}\rangle|^{2}],
\end{align}
where in the RHS $\langle\cdots\rangle$ now denotes a thermal average at fixed disorder and $[\cdots]$ denotes an average over the disorder ensemble. 

\subsection{Phase Diagrams}
$\mathbb{Z}_{N}$ clock models in 2 dimensions have been studied extensively in the literature since they interpolate between the Ising model ($N = 2$) and the XY model ($N\rightarrow \infty$). Notable intermediate cases include the three-state Potts model $(N = 3)$ and the Ashkin-Teller model $(N = 4)$. These models display a rich phase diagram owing to their dihedral symmetry $D_{N} = \mathbb{Z}_{N}\rtimes\mathbb{Z}^{P}_{2}$. In the absence of disorder, it has long been known that the phase diagram of the 2D clock model qualitatively changes with increasing $N$. In particular, for $2\leq N\leq 4$, the $\mathbb{Z}_{N}$ clock model exhibits only two phases: an ordered ferromagnetic phase at low temperature and a disordered paramagnetic phase at high temperature, separated by a single critical point. On the other hand, for $N> 4$, the $\mathbb{Z}_{N}$ clock model exhibits three different phases: an ordered phase, a disordered phase, and an intervening quasi–long-range-ordered (QLRO) phase.~\cite{ZN_QLRO_Existence,ZN_QLRO_Existence_RG,ZN_QLRO_Existence2,ZN_CFT_Analysis,Z6_numerics, ZN_clock_numerics, ZN_clock_numerics2, Domany_1980, Z6_Phase_Diagram, No_Disorder_Clock_model_Phases}. Within this intermediate phase, vortices remain irrelevant and furthermore the $\mathbb{Z}_{N}$ locking terms also become irrelevant, leaving behind an effective Gaussian spin-wave theory with an emergent $U(1)$ symmetry. The existence of the QLRO phase follows from general renormalization group arguments; it does not rely on any fine-tuned choice of the $\beta_m$ couplings. These couplings merely affect nonuniversal properties, such as the precise locations of the two critical points, without altering the qualitative structure of the phase diagram.
\begin{figure*}
    \includegraphics[width=0.95\textwidth]{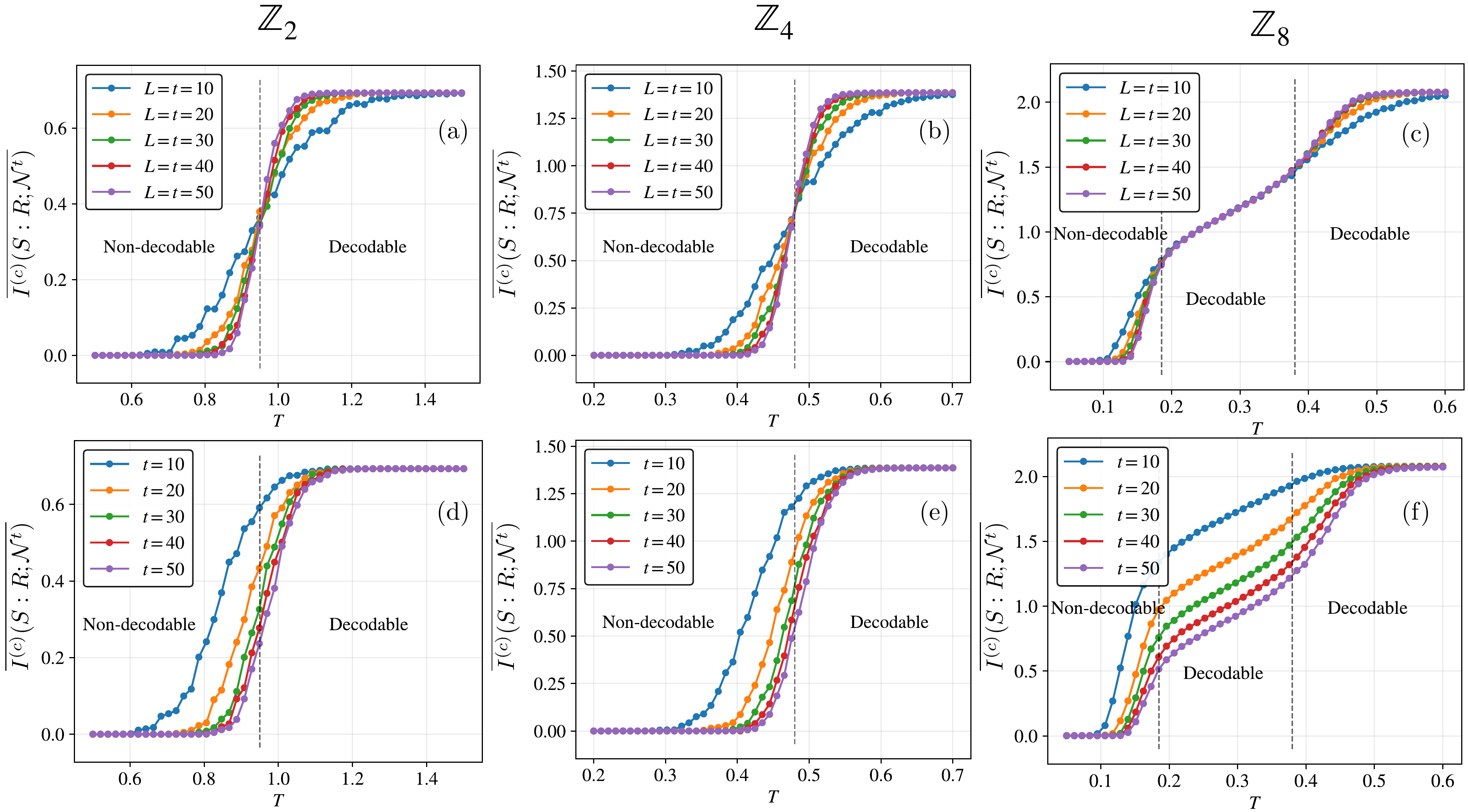}
    \caption{\label{fig:Numerics_CI}  {\bf Numerical results for the coherent information $\overline{I^{(c)}(SM:R;\mathcal{N}^t{})}$:} We compute the measurement-averaged coherent information for $\mathbb{Z}_{2}$ (left), $\mathbb{Z}_{4}$ (middle), and $\mathbb{Z}_{8}$ (right) symmetric circuits subjected to the weak measurement model of~\eqnref{eq:Measurement_model}. To enable a direct comparison across different values of $N$, the vertical axis is plotted in logarithmic units of base $N$. The coherent information is evaluated using the analytic form derived in \eqnref{eq:coherent_information} and sampled numerically via the worm algorithm. Panels (a), (b), and (c) show $\overline{I^{(c)}(SM\!:\!R;\mathcal{N}^{t})}$ along the line $L=t$, while panels (d), (e), and (f) display $\overline{I^{(c)}(SM\!:\!R;\mathcal{N}^{t})}$ for a fixed system size $L=30$ as a function of circuit time $t$. For large $T$ (weak measurements), the encoded quantum information remains robust and the monitored symmetric circuit acts as a stable memory. In contrast, for small $T$ (strong measurements), the information stored in the spin chain is rapidly erased. The coherent information exhibits a sharp transition at the same critical measurement strength as the charge-sharpening transition, confirming that learnability of the global charge coincides with a decodability transition of the emergent quantum error correcting code.}   
\end{figure*}
The fate of this intermediate QLRO phase in the presence of quenched disorder has been heavily debated~\cite{QLRO_Disorder}. It is known that sufficiently strong disorder destroys all signatures of the QLRO phase. However, in our case, we are interested in computing expectation values of the form given in \eqnref{eq:Order_Parameter} for which the disorder distribution $P(\bf{s})$ is effectively proportional to the partition function itself $\mathcal{Z}[\bf{s}]$. This ties the temperature to the disorder and confines us to a special locus known as the \textit{Nishimori line}. The Nishimori line is very special because the disordered model acquires an effective local $\mathbb{Z}_{N}$ symmetry~\cite{Spin_Glasses_Nishimori,YOzeki_1993}. This is entirely unsurprising in our context, since the partition functions we evaluate encode amplitudes of an underlying gauge wavefunction. Along the Nishimori line, it is generally expected that the qualitative structure of the phase diagram matches that of the clean $\mathbb{Z}_{N}$ clock model~\cite{Zn_Gauge_Glass_Nishimori_Analytics,Zn_Gauge_Glass_Nishimori_Numerics}. Which phase the system occupies ultimately depends on the weak measurement channel and the choice of parameters $p_{k}$. Below, we make a concrete choice for these parameters and numerically confirm the emergence of all three phases along the Nishimori line for $N>4$ as the strength of the measurement is varied.

\subsubsection{Numerics}
Here we briefly outline how to compute the quantity $\lAngle  |T_{\hat{\gamma}}|^{2} \rAngle$ given in \eqnref{eq:Order_Parameter}. We consider a one-parameter family of weak measurement channels specified by
\begin{align}\label{eq:Measurement_model}
    p_{k} = \frac{\exp{\frac{1}{T}\cos{\frac{2\pi k}{N}}}}{\mathcal{N}},
\end{align}
where $T$ plays the role of an effective “temperature” and $\mathcal{N}$ is a normalization constant. In the notation of Eq.~\eqref{eq:Hamiltonian}, this corresponds to choosing a single nonzero coupling $\beta_{1} = \frac{1}{T}$ with all other $\beta_{m\neq 1} = 0$. The parameter $T$ controls the strength of the measurement. For small $T$, $p_{k} $ is sharply peaked around $k = 0$ approaching the projective measurement limit. In this regime, measurements profoundly disrupt the state and the global charge sharpens rapidly. On the other hand, for large $T$, $p_{k}\approx \frac{1}{N}$ corresponding to the no measurement limit. 

To proceed, note that the quantity $P(\textbf{s},Q) = \mathcal{Z}[\textbf{s},Q]/(\sum_{Q' = 0}^{N-1}\mathcal{Z}[\textbf{s},Q'])$ represents the probability of occupying homology sector $Q$ in the presence of quenched disorder $\textbf{s}$. These probabilities can be efficiently computed using the worm algorithm~\cite{Worm_Algorithm1,Worm_Algorithm2}, which samples configurations in the high-temperature (current-loop) expansion of the statistical model, as in~\eqnref{eq:partition_function_current_loop}.

Flow configurations obey a divergence-free constraint $\nabla\cdot \mathbf{k} =0$, which the worm algorithm temporarily relaxes by introducing a pair of defect sites, the head and tail of the worm, that violate this constraint. The head is then propagated through local updates weighted by the Boltzmann factor (which depends on the disorder $\mathbf{s}$). When the head returns to the tail, the defects annihilate and the configuration again satisfies all constraints of the original model. Observables are obtained by averaging over these closed-loop configurations, while the intermediate open-worm configurations allow for ergodic sampling of correlation functions. Importantly, the algorithm suffers from minimal critical slowing down and efficiently samples all homology sectors, making it highly efficient for computing the probabilities $P(\textbf{s},Q)$ (see~\cite{Worm_Algorithm1,Worm_Algorithm2} for more details). Once we have $P(\textbf{s},Q)$, we can evaluate $\lAngle  |T_{\hat{\gamma}}|^{2} \rAngle$ by noting that $\lAngle  |T_{\hat{\gamma}}|^{2} \rAngle = [|\sum_{Q = 0}^{N-1}\omega^{Q}P(\textbf{s},Q)|^{2}]$, where $[\cdots]$ denotes the disorder average on the Nishimori line. 

Since our focus is on bulk phase transitions of the statistical mechanical model, we may impose periodic boundary conditions in both lattice directions. The vertical winding sectors can be safely traced over, as they have no physical meaning in the surface code due to the rough boundaries at the top and bottom. In the limit of large lattice sizes, the bulk phase transitions of the statistical model with periodic boundaries coincide with those of the original surface code geometry, since boundary effects become completely negligible. 

In \appref{sec:clock_model}, we analyse the scaling of the partition function $ \mathcal{Z}[\textbf{s};Q]$ with $L$ and $t$ (the dimensions of the lattice) using high and low temperature expansions together with a Gaussian mean-field ansatz appropriate for the QLRO phase. If the net frustration of the phases $\textbf{s}$ along the non-contractible closed curve $\hat{\gamma}$ is given by $Q' = \hat{\gamma}(\textbf{s})$, such that $\mathcal{Z}[\textbf{s};Q']$ contains no topological frustration, then find that $\mathcal{Z}[\textbf{s};Q]$ for any $Q\neq Q'$ scales as:
\begin{align}
    \mathcal{Z}[\textbf{s};Q] \sim \begin{cases}
        \mathcal{Z}[\textbf{s};Q'] + c_{1}(Q-Q')Te^{-\alpha L} &\text{Disordered} \\
        e^{-c_{2}(Q-Q')\frac{t}{L}}\mathcal{Z}[\textbf{s};Q']  &\text{QLRO} \\
        e^{-c_{3}(Q-Q')T}  \mathcal{Z}[\textbf{s};Q'] &\text{Ordered}.
    \end{cases}
\end{align}
This predicts sharpening timescales of the form
\begin{align}
    t_{\#} \sim \begin{cases}
        e^{L}   &\text{Disordered} \\
        L  &\text{QLRO} \\
        \mathcal{O}(1) &\text{Ordered}.
    \end{cases}
\end{align}
We verify these predictions numerically in \figref{fig:Numerics_Var}. 

\subsubsection{Confinement Transition}
We can also consider the case of local charge sharpening. It should be immediate that in the disordered and QLRO phases, local charges never fully sharpen. Using the fact that the sharpening order parameter is precisely the disorder-averaged squared spin–spin correlator (see \eqnref{eq:Local_Sharpening_Order_parameter}), we immediately infer its scaling in the three phases:
\begin{align}
    \lAngle |T_{\hat{\gamma}_{A}}|^{2} \rAngle \sim \begin{cases}
        e^{-cL_{A}} \qquad \text{{Disordered}} \\
        L_{A}^{-2\Delta} \qquad \text{{QLRO}} \\
        \mathcal{O}(1) \qquad \text{{Ordered}}.
    \end{cases}
\end{align}
This behavior admits a transparent interpretation in the gauge-theoretic picture. Recall that the ensemble of post-measurement gauge states is given by
\begin{align}
    \ket{\Phi^{\bf{s}}_{\text{g}}} = \sum_{\substack{\textbf{k}| \nabla\cdot \textbf{k} = 0 \\ \hat{\gamma}(\textbf{k}) = Q}}  \prod_{\mu} \sqrt{p_{k_{\mu};s_{\mu}}}\ket{\bf{k}}.
\end{align}
At large $T$ (low measurement strength), the weights $p_{k_\mu;s_\mu}$ are broad so $\ket{\Phi^{\bf{s}}_{\text{g}}}$ is a coherent sum over essentially all string-net configurations. This is a long range entangled, electrically deconfined state. Equivalently, since the expectation value of the open 'tHooft string decays exponentially $\lAngle |T_{\hat{\gamma}_{A}}|^{2} \rAngle \approx e^{-cL_{A}}$, this phase exhibits linear confinement of magnetic fluxes. 

On the other hand, for very low $T$, measurements become projective and $\ket{\Phi^{\bf{s}}_{\text{g}}}$ approaches a product state $\ket{\bf{s}}$ that is short range entanglemed and electrically confined. Equivalently since $\lAngle |T_{\hat{\gamma}_{A}}|^{2} \rAngle \approx \mathcal{O}(1)$, magnetic fluxes condense in this phase and state is magnetically deconfined. Thus, the ensemble of measured states undergoes a transition from electric deconfinement to electric confinement. In the intermediate QLRO phase, the 't~Hooft operator decays as a power law, $\lAngle |T_{\hat{\gamma}_{A}}|^{2} \rAngle \approx e^{-cL_{A}}$, indicating that the state supports gapless modes. This corresponds to a Coulomb phase with emergent photons. Recall that in this regime the partition function is governed by an effective Gaussian spin-wave theory, whose long-wavelength excitations capture smooth fluctuations of the coarse-grained ``height" field. Since the height variables $\theta_p$ encode the magnetic flux number, the emergent photons can be understood as collective, long-wavelength oscillations of local magnetic fluxes.

\subsection{Decodability Transition}
Upto this point, we have interpreted the sharpening transition as a \textit{learnability transition}, a change in how quickly local charge measurements reveal the global charge of the spin chain. We now recast the same phenomenon as a \textit{decodability transition}, namely a change in how rapidly quantum information encoded in terms of the logical spin states (\eqnref{eq:logical_subspace}) is destroyed by the monitored symmetric circuit. In the charge-fuzzy phase, the classical information corresponding to the global charge of the spin chain remains hidden upto timescales that are extensive in system size. In this phase, the monitored symmetric circuit functions as an effective quantum memory, and the logical qudit encoded in the initial state remains protected over these extensive timescales. By contrast, in the charge-sharp phase, the global charge is revealed rapidly, and the logical information stored in the initial spin state is quickly erased. 

We now consider the coherent information of the spin chain evolving under the combined unitary+measurement channel. A key distinction in this setting is that the measurement record constitutes classical information and must be explicitly retained. Consequently, the effective quantum channel that acts on the spin chain $S$ is really given by
\begin{align}
    \mathcal{N}^{t}(\rho_{S}) = \sum_{\textbf{s}} {\mathcal{N}^{t}}_{s}\rho_{S}\mathcal{N}^{t\dagger}_{\textbf{s}}\otimes |s\rangle\langle s|_{M},
\end{align}
where $\mathcal{N}^{t}_{\textbf{s}}$ denotes the monitored circuit evolution operator at time $t$, conditioned on a specific trajectory of weak measurement outcomes $\mathbf{s}$, and $|\mathbf{s}\rangle\langle \mathbf{s}|_{M}$ denotes the state of the classical measurement register $M$ that stores the full measurement record.

We introduce a reference system $R$ that is maximally entangled with the logical states of the spin chain as follows
\begin{align}
    \ket{\Psi_{i}}_{SR} = \frac{1}{\sqrt{N}}\sum_{Q=0}^{N-1}\ket{\psi(Q)}_{S}\otimes \ket{Q}_{R}.
\end{align}
Under the quantum channel, this state evolves to 
\begin{align}
    \rho_{SRM} = \sum_{\textbf{s}} {\mathcal{N}^{t}}_{s}|\Psi_{i}\rangle_{SR} \langle\Psi_{i}|_{SR}\mathcal{N}^{t\dagger}_{\textbf{s}}\otimes |s\rangle\langle s|_{M}.
\end{align}
The coherent information, conditioned on access to the measurement outcomes, is then given by
\begin{align}
    I^{(c)}(SM:R;\mathcal{N}^{t}) = S(\mathcal{N}^{t}(\rho_{SM})) -  S(\mathcal{N}^{t}(\rho_{SRM}))
\end{align}
Again, averaging this quantity over random multiplicity tensors in the large bond dimension limit allows us to express this quantity as the coherent information of the bulk surface code (see \secref{sec:entropy_equality})
\begin{align}
    \overline{I^{(c)}(SM:R;\mathcal{N}^{t})} = S(\mathcal{E}(\rho_{gM})) -  S(\mathcal{E}(\rho_{gRM}))
\end{align}
where $\mathcal{E}$ now only contains measurement channels and not the unitary evolution. The initial state of the reference $R$ and the gauge system $g$ is given by
\begin{align}
    \ket{\Phi_{\text{g}R}} = \frac{1}{\sqrt{N}}\sum_{Q=0}^{N-1}\ket{\Phi_{g}(Q)}\otimes \ket{Q}_{R} 
\end{align}
Recall again that $\ket{\Phi_{g}(Q)}$ are just the surface code logical states. We can now compute the above entropies in the bulk surface code and show that it undergoes a transition. In terms of the partition functions $\mathcal{Z}[\textbf{s},Q] =\braket{\Phi^{\boldsymbol{s}}_{\text{g}}(Q)}{\Phi^{\boldsymbol{s}}_{\text{g}}(Q)}$, it is easily shown that 
\begin{align}
    S(\mathcal{E}(\rho_{gMR})) &= -\sum_{\textbf{s}}\mathcal{Z}[\textbf{s}]\log \mathcal{Z}[\textbf{s}] \\
    S(\mathcal{E}(\rho_{gM})) &= -\sum_{\textbf{s},Q}\frac{\mathcal{Z}[\textbf{s},Q]}{N}\log \frac{\mathcal{Z}[\textbf{s},Q]}{N}
\end{align}
where $\mathcal{Z}[\textbf{s}] =\frac{1}{N}\sum_{Q}\mathcal{Z}[\textbf{s},Q] $ and for convenience we've assumed the normalization $\frac{1}{N}\sum_{\textbf{s},Q}\mathcal{Z}[\textbf{s},Q] = 1$. This immediately implies that 
\begin{align}\label{eq:coherent_information}
    \overline{I^{(c)}(SM:R;\mathcal{N}^{t})} &= -\sum_{\textbf{s},Q}\frac{\mathcal{Z}[\textbf{s},Q]}{N}\log \frac{\mathcal{Z}[\textbf{s},Q]}{\sum_{Q'}\mathcal{Z}[\textbf{s},Q']} \nonumber \\
    &= \lAngle S(\rho_{gM}|\textbf{s})\rAngle
\end{align}
where $S(\rho_{gM}|\textbf{s})$ is just the Von Neumann entropy of the state $\rho_{gM}$ conditioned on measurement outcome $\textbf{s}$. Similar calculations of the coherent information have been carried out for decohered topological codes in~\cite{Fan_Bao_Altman,Lee2023, Lee_Exact_Diagonalization,Lee_CSS_Codes,Error_Threshold_General_Stat_Mech_Mappings,Intrinsic_Error_Thresholds,Tapestry_of_dualities,Vijay2025_appear}. In \figref{fig:Numerics_CI}, we use the worm algorithm to numerically evaluate the coherent information and find that it exhibits a sharp transition at exactly the same critical measurement strength as the charge-sharpening transition. This confirms that the learnability of the global charge coincides with the loss of quantum information encoded in the surface-code logical states.

\section{Discussion}

In this work, we introduced a holographic framework for studying the dynamical phases of symmetric quantum circuits. We have shown that a local quantum circuit endowed with a global $G$-symmetry admits a bulk-boundary correspondence whereby the final state of the spin chain can be expressed as a contraction of bulk multiplicity tensors and a $G$-gauge wavefunction living in one higher dimension. In the large-multiplicity limit, averages of a broad class of charge observables map directly to corresponding gauge-theoretic observables. In the case of unitary dynamics, we further demonstrated that the resulting gauge wavefunction becomes a coherent superposition of all allowed string-net configurations, signalling long-range entanglement.

For $\mathbb{Z}_N$-symmetric circuits composed of symmetric unitary gates interspersed with local symmetry-respecting noise, the bulk realizes a noisy $\mathbb{Z}_N$ surface code. This observation naturally identifies a set of logical spin states, one in each global charge sector, which inherit the topological protection of the bulk surface code. We made this connection explicit by showing that the coherent information of the spin chain, averaged over random gates in the large-multiplicity limit,  coincides with the coherent information of the bulk $\mathbb{Z}_N$ surface code, which is intrinsically robust against such local noise.



Finally, we analyzed $\mathbb{Z}_N$-symmetric, monitored circuits subject to weak charge measurements at each timestep. In the bulk description, this corresponds to a $\mathbb{Z}_{N}$ surface code undergoing weak measurements on each link. We show that the point at which the observer acquires classical information about the global charge coincides with the point at which measurements destroy the underlying quantum information encoded in the bulk code. Moreover, we showed that beyond the charge-sharpening point, the measured gauge wavefunction effectively undergoes a confinement transition. For $N \le 4$ there is a single transition from a deconfined phase with $t_{\#} \sim e^{L}$ to a confined phase with $t_{\#} \sim \mathcal{O}(1)$, while for $N > 4$ measurements generically produce an intermediate phase with $t_{\#} \sim L$, mirroring the scaling behavior found in $U(1)$-symmetric circuits. In this regime, the bulk gauge wavefunction realizes a Coulomb phase with emergent gapless photons.


While our analysis has focused on dynamical phases in Abelian symmetric circuits, it extends naturally to non-Abelian symmetries. An important direction for future work is to investigate the robustness of the error-correcting properties identified here in that setting, and to determine whether non-Abelian circuits support genuinely new measurement or decoherence driven phases with no Abelian analogue.

Furthermore, as discussed earlier, the charge sharpening transition can be viewed as the restoration of strong symmetry from an initial SWSSB phase, which is closely tied to the notion of charge scrambling~\cite{Lee2026}. Indeed, for U(1) symmetric systems, it was shown that the sharp phase is incompatible with SWSSB~\cite{singh2025mixedstatelearnabilitytransitionsmonitored}.
Therefore, we expect the holographic framework developed here to be useful for analyzing entanglement and operator dynamics in symmetric quantum circuits, and for addressing questions of thermalization and scrambling in the presence of conserved charges.


\vspace{5pt} 

\emph{Notes added}: Near the completion, we became aware of an independent work that is broadly related~\cite{Yujie2025}. We also draw attention for the forthcoming work on monitored symmetric circuits using higher Lindbladians~\cite{Makinde2025}.

\acknowledgments
We thank Romain Vasseur, Olumakinde Ogunnaike, Rishi Lohar, and Michael Levin for fruitful discussions and comments. A.V. thanks Zack Weinstein for directing us to the Worm algorithm that significantly expedited the numerical simulation.
The work is supported by the faculty startup grant at the University of Illinois, Urbana-Champaign and the IBM-Illinois Discovery
Accelerator Institute.

\bibliography{refs}

\newpage
\appendix

\section{Averaging over random multiplicity vectors $\ket{T_{v}}$}\label{sec:Average_Multiplicity}
Here we provide additional details on the averaging over random multiplicity tensors. Let the on-site Hilbert space decompose into irreducible representations as,
\begin{align}
    \mathcal{H} = \bigoplus_{j} n\mathcal{R}_{j},
\end{align}
where $R_{j}$ are assumed to be one dimensional Abelian irreps. The multiplicity vertex tensors are given by 
\begin{align}
    \ket{T_{v}} = \sum_{\bm{j}, \bm{a}} (T_{v})^{\bm{j}}_{\bm{a}}  |\bm{j}, \bm{a} \rangle    .
\end{align}
The tensor components $(T_{v})^{\bm{j}}_{\bm{a}}$ are taken to be independent Gaussian random variables with variances $1/D_{J_{\text{in}}}$ (or $1/D_{J_{\text{out}}}$)~\footnote{Since we eventually project onto the $J_{\text{in}} = J_{\text{out}}$ sector, these choices are equivalent}. Our objective is to compute ensemble averages of the moments 
\begin{align}
    \overline{\ket{T_{v}}\bra{T_{v}}^{\otimes k}}.
\end{align}
First consider the case when $k = 2$. Expanding this out in the above basis, we obtain 
\begin{align}
   \overline{\ket{T_{v}}\bra{T_{v}}^{\otimes 2}} =  \sum_{\substack{\bm{j}_{1},\bm{j}_{2}\bm{j}_{3}\bm{j}_{4} \\ \bm{a}_{1},\bm{a}_{2}\bm{a}_{3}\bm{a}_{4}}} \overline{(T_{v})^{\bm{j}_{1}}_{\bm{a}_{1}}(T_{v})^{\bm{j}_{2}}_{\bm{a}_{2}} (T^{*}_{v})^{\bm{j}_{3}}_{\bm{a}_{3}}(T^{*}_{v})^{\bm{j}_{4}}_{\bm{a}_{4}}}     \nonumber \\
    |\bm{j}_{1}, \bm{a}_{1} \rangle    |\bm{j}_{2}, \bm{a}_{2} \rangle    \langle \bm{j}_{3}, \bm{a}_{3}    |\langle   \bm{j}_{4}, \bm{a}_{4} | .
\end{align}
Using Wick's theorem, the above average breaks up into a sum of 2 terms which are
\begin{align}
    \frac{1}{D_{J_{1;\text{in}}}D_{J_{2;\text{in}}}}\bigg(\delta^{\bm{j}_{1}\bm{j}_{3}}\delta^{\bm{j}_{2}\bm{j}_{4}} \delta_{\bm{a}_{1}\bm{a}_{3}}\delta_{\bm{a}_{2}\bm{a}_{4}} + \delta^{\bm{j}_{1}\bm{j}_{4}}\delta^{\bm{j}_{2}\bm{j}_{3}} \delta_{\bm{a}_{1}\bm{a}_{4}}\delta_{\bm{a}_{2}\bm{a}_{3}}  \bigg) \nonumber.
\end{align}
Note the appearance of total incoming-charge dependent variances. If $D_{J}$ is the same for all $J$, we can move this prefactor out of the sum and express the result as a simple sum of the identity and swap operator. However, since we cannot generally do this, we define the following operator
\begin{align}
    M_v &= \sum_{\bm{j}, \bm{a}} \frac{1}{\sqrt{D_{J_{\text{in}}}}} |\bm{j}, \bm{a} \rangle\langle \bm{j}, \bm{a}    | .
\end{align}
Now note that
\begin{align}
    M_v^{\otimes 2}P_{e} M_v^{\otimes 2} &= \sum_{\substack{\bm{j}_{1},\bm{j}_{2} \\ \bm{a}_{1},\bm{a}_{2}}} \frac{1}{D_{J_{1;\text{in}}}D_{J_{2;\text{in}}}} \nonumber \\
    &\times|\bm{j}_{1}, \bm{a}_{1} \rangle    |\bm{j}_{2}, \bm{a}_{2} \rangle    \langle \bm{j}_{1}, \bm{a}_{1}    |\langle   \bm{j}_{2}, \bm{a}_{2} |   \\
     M_v^{\otimes 2}P_{\tau_{2}} M_v^{\otimes 2} &= \sum_{\substack{\bm{j}_{1},\bm{j}_{2} \\ \bm{a}_{1},\bm{a}_{2}}} \frac{1}{D_{J_{1;\text{in}}}D_{J_{2;\text{in}}}} \nonumber \\
     &\times|\bm{j}_{1}, \bm{a}_{1} \rangle    |\bm{j}_{2}, \bm{a}_{2} \rangle    \langle \bm{j}_{2}, \bm{a}_{2}    |\langle   \bm{j}_{1}, \bm{a}_{1} |,
\end{align}
where $P_{e}$ and $P_{\tau_{2}}$ are the identity and swap operators respectively. It is then clear that
\begin{align}
    \overline{|T_{v}\rangle \langle T_{v}|^{\otimes 2}} =  M^{\otimes 2}P_{e} M^{\otimes 2} +  M^{\otimes 2}P_{\tau_{2}} M^{\otimes 2}.
\end{align}
Furthermore, it is easily verified that this pattern continues to hold for general $k$
\begin{align}\label{eq:average_multiplicity_tensors}
    \overline{|T_{v}\rangle \langle T_{v}|^{\otimes k}} =  \sum_{\sigma \in S_{k}}M^{\otimes k}P_{\sigma} M^{\otimes k}.
\end{align}
The net effect of this averaging is to dress gauge wavefunction as \( \ket{\Phi_{g}} \mapsto \prod_{v}M_{v} \ket{\Phi_{g}}\).


\section{Unitarity of non-abelian $G$-symmetric gates}
\label{app:Unitarity_nonabelian_gates}
Recall that a $G$-symmetric two-site gate admits the tensor decomposition,
\begin{align}
    A^{J_{1}J_{2}J_{3}J_{4}} = \sum_{e,\mu_{e}}  \big(C^{j_{1}j_{2}e}_{\mu_{1}\mu_{2}\mu_{e}}\big)^{*}C^{j_{3}j_{4}e}_{\mu_{3}\mu_{4}\mu_{e}}T^{j_{1}j_{2}j_{3}j_{4}; e}_{a_{1}a_{2}a_{3}a_{4}},
\end{align}
where $J = (j,\mu,a)$. This decomposition is block-diagonal in the intermediate (fusion) charge $e$. Unitarity of the gate requires
\begin{align}\label{eq:Unitarity_condition}
    \sum_{J_{3}J_{4}}A^{J_{1}J_{2}J_{3}J_{4}} (A^{J_{5}J_{6}J_{3}J_{4}})^{*} = \delta^{J_{1}J_{5}}\delta^{J_{2}J_{6}}.
\end{align}
Expanding this out, we obtain 
\begin{align}
   \delta^{J_{1}J_{5}}\delta^{J_{2}J_{6}} =  &\sum_{J_{3}J_{4}}\sum_{e,\mu_{e}}  \big(C^{j_{1}j_{2}e}_{\mu_{1}\mu_{2}\mu_{e}}\big)^{*}C^{j_{3}j_{4}e}_{\mu_{3}\mu_{4}\mu_{e}}T^{j_{1}j_{2}j_{3}j_{4}; e}_{a_{1}a_{2}a_{3}a_{4}} \nonumber \\
   \times &\sum_{f,\mu_{f}}  C^{j_{5}j_{6}f}_{\mu_{5}\mu_{6}\mu_{f}} \big(C^{j_{3}j_{4}f}_{\mu_{3}\mu_{4}\mu_{f}}\big)^{*}(T^{j_{5}j_{6}j_{3}j_{4}; f}_{a_{5}a_{6}a_{3}a_{4}})^{*}.
\end{align}
Then, using the normalization convention for the Clebsch-Gordon coefficients in \eqnref{eq:CS_normalization}, we obtain 
\begin{align}
   \delta^{J_{1}J_{5}}\delta^{J_{2}J_{6}} =  &\sum_{e,\mu_{e}}  \big(C^{j_{1}j_{2}e}_{\mu_{1}\mu_{2}\mu_{e}}\big)^{*}  C^{j_{5}j_{6}e}_{\mu_{5}\mu_{6}\mu_{e}}  \nonumber \\
   \times & \sum_{j_{3}j_{4}}\sum_{a_{3}a_{4}} T^{j_{1}j_{2}j_{3}j_{4}; e}_{a_{1}a_{2}a_{3}a_{4}}  (T^{j_{5}j_{6}j_{3}j_{4}; e}_{a_{5}a_{6}a_{3}a_{4}})^{*}.
\end{align}
If we assume that 
\begin{align}
    \sum_{j_{3}j_{4}}\sum_{a_{3}a_{4}} T^{j_{1}j_{2}j_{3}j_{4}; e}_{a_{1}a_{2}a_{3}a_{4}}  (T^{j_{5}j_{6}j_{3}j_{4}; e}_{a_{5}a_{6}a_{3}a_{4}})^{*} = \delta^{j_{1}j_{5}}\delta^{j_{2}j_{6}}\delta_{a_{1}a_{5}}\delta_{a_{2}a_{6}},
\end{align}
namely if $T$ is unitary within each block labeled by the total charge $e$, when viewed as a matrix in the combined charge and multiplicity indices, then \eqnref{eq:Unitarity_condition} holds for the full gate.

\section{Boundary entropy = Bulk entropy}\label{sec:entropy_equality}
The bulk--boundary correspondence in the general mixed-state setting takes the form
\begin{align}
    \rho_{\mathrm{Bdy}}
    =
    \Bigl( \bigotimes_{v} \!\bra{T_{v}} \Bigr)\,
        E\, \rho_{\mathrm{g}}\, E^{\dagger}\,
    \Bigl( \bigotimes_{v} \!\ket{T_{v}} \Bigr),
\end{align}
where $\rho_{\mathrm{g}}$ is the bulk gauge density matrix and $E$ denotes the extended-Hilbert-space embedding map. We assume that the initial state of the spin chain is pure.

Our goal is to show that, in the limit of \emph{large multiplicity}, the entropy of the boundary state matches that of the bulk state:
\begin{align}
    \overline{S(\rho_{\mathrm{Bdy}})}
    = S(\rho_{\mathrm{g}}).
\end{align}
To establish this identity, it suffices to demonstrate equality for all Rényi entropies:
\begin{align}\label{eq:Renyis_equal}
    \overline{S^{(n)}(\rho_{\mathrm{Bdy}})}
    = S^{(n)}(\rho_{\mathrm{g}}),
    \qquad n \in \mathbb{N},
\end{align}
since the von Neumann entropy is obtained via an analytic continuation of the Renyi entropies,
$S(\rho) = \lim_{n\to 1} S^{(n)}(\rho)$. 
A key property of random multiplicity tensors in the large bond-dimension limit is the self-averaging identity~\cite{RTNs_2016,qi2022emergentbulkgaugefield}
\begin{align}
    \overline{\bigg[\ln\frac{\Tr \rho_{\text{Bdy}}^{n}}{(\Tr \rho_{\text{Bdy}})^{n}}\bigg]} = \frac{\overline{\Tr \rho_{\text{Bdy}}^{n}}}{\overline{(\Tr \rho_{\text{Bdy}})^{n}}},
\end{align}
which holds because fluctuations of the vertex tensors are suppressed by powers of the multiplicity dimension.
Let us first write the $n$-th normalized moment of $\rho_{\text{Bdy}}$ as  
\begin{align}
    \frac{\Tr \rho_{\text{Bdy}}^{n}}{(\Tr \rho_{\text{Bdy}})^{n}} = \frac{\Tr \rho^{\otimes n}_{\text{Bdy}} P_{\tau _{n}}}{(\Tr \rho^{\otimes n}_{\text{Bdy}} P_{e})},
\end{align}
where $P_{\tau _{n}}$ denotes the cyclic permutation on $n$ copies of the state. Using the bulk-boundary correspondence, the latter can be written as 
\begin{align}
    \frac{\Tr \rho^{\otimes n}_{\text{Bdy}} P_{\tau _{n}}}{(\Tr \rho^{\otimes n}_{\text{Bdy}} P_e)} = \frac{\Tr
      \rho_{\text{Bulk}}^{\otimes n} P_{\tau _{n}}
      \Big( \bigotimes_{v} \ket{T_{v}}\bra{T_{v}}^{\otimes n} \Big)}{\Tr
      \rho_{\text{Bulk}}^{\otimes n}P_e
      \Big( \bigotimes_{v} \ket{T_{v}}\bra{T_{v}}^{\otimes n} \Big)},
\end{align}
where for convenience, we've introduced notation $\rho_{\text{Bulk}}= E \rho_{g} E^{\dagger}$. We now consider the averaged numerator and denominator separately. Using~\eqnref{eq:average_multiplicity_tensors},
the averaged numerator becomes
\begin{align}
        \Tr
      {\rho}_{\text{Bulk}}^{\otimes n} P_{\tau _{n}}&\overline{
      \Big( \bigotimes_{v} \ket{T_{v}}\bra{T_{v}}^{\otimes n} \Big)} \nonumber \\
      &=\sum_{\{ \sigma({x})\}}  \Tr{
      \tilde{\rho}_{\text{Bulk}}^{\otimes n} P_{\tau _{n}}\prod_{x}P_{\sigma({x})}} \nonumber \\
      &=\sum_{\{ \sigma({x})\}}  \Tr{
      \tilde{\rho}_{\text{g}}^{\otimes n} E^{\dagger}P_{\tau _{n}}\prod_{x}P_{\sigma({x})} E}.
\end{align}
The product runs over all vertices $x$ and $\tilde{\rho}_{\text{g}} = \prod_{x}M_x{\rho_{g}} \prod_{x}M_x$ is the dressed gauge state. 

The presence of the canonical isometry $E$ is crucial here. Recall that $E$
factorizes over all links of the lattice,
\begin{align}
    E = \prod_{\langle x y \rangle} E_{xy}.
\end{align}
We now show that any contribution in which neighboring permutations do not align is suppressed by powers of the multiplicity network bond dimension. For simplicity, let us focus on the case $n=2$, corresponding to the
computation of the purity. In this case, each local permutation $\sigma(x)$ can be either the identity $e$ or the swap $\tau$.

Let $E_{xy}$ denote the isometry associated with the link from vertex $x$ to vertex $y$, which we write as  
\begin{align}
    E_{xy} = \sum_{j}\frac{1}{\sqrt{n_{j}}}\sum_{a}\ket{j^{*},a}_{xy}\ket{j,a}_{yx}\bra{j},
\end{align}
where $\ket{\cdot}_{xy}$ belongs to the Hilbert space at vertex $x$ and $\ket{\cdot}_{yx}$ to the Hilbert space at vertex $y$, and $n_{j}$ is the multiplicity of irrep $j$. If $\sigma(x) =  \sigma(y) = e$, then $E^{\otimes 2 \dagger}_{xy}\sigma(x)\sigma(y) E_{xy}^{\otimes 2} = \mathbb{I}$ whereas if $\sigma(x) =  \sigma(y)= \tau$, then $E^{\otimes 2 \dagger}_{xy}\sigma(x) \sigma(y) E_{xy}^{\otimes 2} = \tau({xy})$ where $\tau({xy})$ is the swap operator acting on the two replicas of the \emph{link} degree of freedom $(xy)$. To see this, note that $E^{\otimes 2 \dagger}_{xy}\tau(x)\tau(y) E_{xy}^{\otimes 2}$ is just given by 
\begin{align}
    &\sum_{\substack{j_{1}j_{2}\\a_{1},a_{2}}}\frac{1}{\sqrt{n_{j_{1}}n_{j_{2}}}}\ket{j_{1}j_{2}}\bra{j_{1}^{*},a_{1};j^{*}_{2},a_{2}}_{xy}\bra{j_{1},a_{1};j_{2},a_{2}}_{yx} \times \nonumber \\
    &\sum_{\substack{k_{1}k_{2}\\b_{1},b_{2}}}\frac{1}{\sqrt{n_{k_{1}}n_{k_{2}}}}\ket{k_{2}^{*},b_{2};k^{*}_{1},b_{1}}_{xy}\ket{k_{2},b_{2};k_{1},b_{1}}_{yx}\bra{k_{1}k_{2}} \nonumber \\
    = &\sum_{\substack{j_{1}j_{2}k_{1}k_{2}\\a_{1},a_{2}b_{1}b_{2}}}\frac{1}{\sqrt{n_{j_{1}}n_{j_{2}}n_{k_{1}}n_{k_{2}}}}\ket{j_{1}j_{2}}\bra{k_{1}k_{2}} \times \nonumber \\
    & \qquad \qquad  \delta_{j_{1}k_{2}}\delta_{a_{1}b_{2}}\delta_{j_{2}k_{1}}\delta_{a_{2}b_{1}}\delta_{j_{1}k_{2}}\delta_{a_{1}b_{2}}\delta_{j_{2}k_{1}}\delta_{a_{2}b_{1}} \nonumber \\
    &= \sum_{j,k}\ket{j,k}\bra{k,j} = \tau({xy}).
\end{align}
On the other hand if $\sigma(x) = \tau$ but $\sigma(y) = e$, the same calculation yields instead
\begin{align}
    &\sum_{\substack{j_{1}j_{2}\\a_{1},a_{2}}}\frac{1}{\sqrt{n_{j_{1}}n_{j_{2}}}}\ket{j_{1}j_{2}}\bra{j_{1}^{*},a_{1};j^{*}_{2},a_{2}}_{xy}\bra{j_{1},a_{1};j_{2},a_{2}}_{yx} \times \nonumber \\
    &\sum_{\substack{k_{1}k_{2}\\b_{1},b_{2}}}\frac{1}{\sqrt{n_{k_{1}}n_{k_{2}}}}\ket{k_{2}^{*},b_{2};k^{*}_{1},b_{1}}_{xy}\ket{k_{1},b_{1};k_{2},b_{2}}_{yx}\bra{k_{1}k_{2}} \nonumber \\
    = &\sum_{\substack{j_{1}j_{2}k_{1}k_{2}\\a_{1},a_{2}b_{1}b_{2}}}\frac{1}{\sqrt{n_{j_{1}}n_{j_{2}}n_{k_{1}}n_{k_{2}}}}\ket{j_{1}j_{2}}\bra{k_{1}k_{2}} \times \nonumber \\
    & \qquad \qquad  \delta_{j_{1}k_{2}}\delta_{a_{1}b_{2}}\delta_{j_{2}k_{1}}\delta_{a_{2}b_{1}}\delta_{j_{1}k_{1}}\delta_{a_{1}b_{1}}\delta_{j_{2}k_{2}}\delta_{a_{2}b_{2}} \nonumber \\
    &= \sum_{j}\frac{1}{{n_{j}}}\ket{j,j}\bra{j,j}.
\end{align}
The operator $E^{\otimes 2 \,\dagger}\,\tau({x})\,E^{\otimes 2}$ freezes the same charge across all the replicas of the link. But crucially note the additional factor of $\frac{1}{{n_{j}}}$ outfront. In the
limit of large multiplicities $n_{j}\to\infty$, such contributions become strongly suppressed. This observation extends to arbitrary replica number $n>2$. Whenever the neighboring permutations $\sigma({x})$ and $\sigma({y})$ differ, i.e. whenever $\sigma({x})\sigma({y})^{-1}$ contains a nontrivial cycle, the resulting contraction introduces additional suppressing factors in multiplicity. Consequently, all configurations with locally misaligned permutations are parametrically suppressed in the large-multiplicity
limit.

Thus, the dominant contribution always stems from the configuration in which \emph{all} local permutations are aligned. This is an expression of \textit{replica symmetry breaking} in the large-multiplicity limit. The specific choice of the frozen permutation pattern is not arbitrary: it is fixed by the boundary conditions, which in turn are determined by the observable we are computing. In this discussion we restrict to the case in
which a single permutation is imposed uniformly along the boundary. This simplifies the analysis considerably. When the boundary permutation is $\tau_{n}$, all bulk permutations are forced into $\tau_{n}$; when the boundary permutation is $e$, all bulk permutations are forced into $e$.

This implies that, after averaging over multiplicity tensors, the normalized replica moments become:
\begin{align}
    \Tr
      {\rho_{\text{Bulk}}^{\otimes n} \tau _{n}\overline{
      \Big( \bigotimes_{v} \ket{T_{v}}\bra{T_{v}}^{\otimes n} \Big)}} &\approx \Tr{\tilde{\rho}^{n}_{\text{g}}} \\
      \Tr
      {\rho_{\text{Bulk}}^{\otimes n} e\overline{
      \Big( \bigotimes_{v} \ket{T_{v}}\bra{T_{v}}^{\otimes n} \Big)}} &\approx \Tr{\tilde{\rho}_{\text{g}} }^{n}.
\end{align}
In the $\mathbb{Z}_{N}$ case, dressing the gauge state with the operator $M$ has no effect, so \eqnref{eq:Renyis_equal} follows immediately.

Finally, suppose that an external reference system $R$ is entangled with the charge degrees of freedom of the initial spin-chain state but entirely unentangled with multiplicity degrees of freedom. In this case the
bulk--boundary correspondence for the final boundary state on $S$ together with the reference $R$ is trivially modified to
\begin{align}
    \rho_{SR}
    =
    \Bigl( \bigotimes_{v} \!\bra{T_{v}} \Bigr)\,
        E\, \rho_{\mathrm{gR}}\, E^{\dagger}\,
    \Bigl( \bigotimes_{v} \!\ket{T_{v}} \Bigr),
\end{align}
where $\rho_{\mathrm{gR}}$ denotes the joint mixed state of the gauge theory and the reference system. The previous analysis now applies verbatim and the replica calculation under this modified correspondence immediately yields
\begin{align}
    \overline{S(\rho_{SR})}
    =
    S(\rho_{\mathrm{gR}}).
\end{align}

\section{Charge Sharpening Diagnostic}\label{sec:Sharpening_Diagnostic}
Our goal in this section is to prove Eq.~\eqref{eq:averaged_variance} of the
main text. The calculation closely parallels the entropy argument presented in~\secref{sec:entropy_equality}. Recall that the diagnostic for charge sharpening is the quantity 
\begin{align}\label{eq:Edwards_Anderson_Type_OP}
    \overline{\lAngle |\langle Z_{\text{tot}}\rangle_{\mathbf{s},t} |^{2}\rAngle}  
    &= \overline{\bigg[\sum_{\textbf{s}}\frac{\bra{\psi^{\bf{s}}(t)}^{\otimes 2} Z_{\text{tot}}^{\dagger}\otimes  Z_{\text{tot}}    \ket{\psi^{\bf{s}}(t)}^{\otimes 2}}{\braket{\psi^{\bf{s}}(t)}{\psi^{\bf{s}}(t)}}\bigg]} \nonumber.
\end{align}
Again, using the bulk-boundary correspondence, the RHS becomes
\begin{align}
   \overline{ \Bigg[\frac{\bra{\Phi^{\boldsymbol{s}}_\text{g}}^{\otimes 2} E^{\otimes 2 \dagger} \bigotimes_{v}\ket{T_v} \bra{T_v}^{\otimes 2} Z_{\text{tot}}^{\dagger}\otimes  Z_{\text{tot}} E^{\otimes 2 }   \ket{\Phi^{\boldsymbol{s}}_\text{g}}^{\otimes 2}}{\bra{\Phi^{\boldsymbol{s}}_\text{g}} E \bigotimes_{v}\ket{T_v} \bra{T_v}  E  \ket{\Phi^{\boldsymbol{s}}_\text{g}}} \Bigg]}, \nonumber
\end{align} 
where $\ket{\Phi^{\mathbf{s}}_{\mathrm{g}}}$ is the bulk gauge wavefunction associated with conditioned measurement record $\mathbf{s}$.

At large multiplicities, the Gaussian averages factorize over both the numerator and the denominator, allowing each to be computed independently. Since we specialize here to $\mathbb{Z}_{N}$ gauge theory, we may ignore all multiplicity-dressing factors $M$. Under this averaging, the denominator reduces immediately to $\braket{\Phi^{\bf{s}}_{g}}{\Phi^{\bf{s}}_{g}}$. On the other hand, averaging over $\ket{T_v}\bra{T_v}^{\otimes 2}$ yields a sum over the identity $e$ and swap operator $\tau$ at each vertex $x$. Thus, we can find that 
\begin{align}
    &\overline{\lAngle |\langle Z_{\text{tot}}\rangle_{\mathbf{s},t} |^{2}\rAngle} =\nonumber \\
    &\frac{\sum_{\Omega \subseteq V} \bra{\Phi^{\boldsymbol{s}}_{\text{g}}}^{\otimes 2}E^{\otimes 2 \dagger}\tau({\Omega}) Z_{\text{tot}}^{\dagger}\otimes  Z_{\text{tot}} E^{\otimes 2 }\ket{\Phi^{\boldsymbol{s}}_{\text{g}}}^{\otimes 2}}{\braket{\Phi^{\bf{s}}_{g}}{\Phi^{\bf{s}}_{g}}},
\end{align}
where $\tau({\Omega}) = \prod_{x\in\Omega} \tau({x})$ and the sum is over all disconnected subsets $\Omega \subseteq V$ of the two-dimensional spacetime lattice. To further proceed, we once again invoke the fact that $n_j \gg 1$. Using the same argument as  in~\secref{sec:entropy_equality}, we conclude that the numerator is completely dominated by the saddle point $\Omega = \phi$  where all permutations are $e$. Again this is an expression of replica symmetry breaking in the volume law phase We thus obtain
\begin{align}
   \overline{\lAngle |\langle Z_{\text{tot}}\rangle_{\mathbf{s},t} |^{2}\rAngle}  &= \sum_{\bf{s}}p({\bf{s}})\Bigg|\frac{\bra{\Phi^{\boldsymbol{s}}_{\text{g}}} {T_{\hat{\gamma}}}\ket{\Phi^{\boldsymbol{s}}_{\text{g}}}}{\braket{\Phi^{\boldsymbol{s}}_{\text{g}}}{\Phi^{\boldsymbol{s}}_{\text{g}}}}\Bigg|^{2}  \nonumber \\
    &= \lAngle |T_{\hat{\gamma}}|^{2} \rAngle,
\end{align}
where $p(\textbf{s}) = \braket{\Phi^{\textbf{s}}_{g}}{\Phi^{\textbf{s}}_{g}}$ and $T_{\hat{\gamma}} = E^{\dagger} Z_{\text{tot}} E $ is the 't Hooft loop inserted at the final timeslice  boundary.

\section{Relative Entropy}\label{sec:relative_entropy}
Finally, before proceeding, we consider an additional diagnostic for the robustness of the logical qudit encoded in spin states $\{ \ket{\psi(Q)}\}$. This is the quantum relative entropy between two distinct logical states following the action of $\mathcal{N}^{t}$. 

Recall that for two density matrices $\rho$ and $\sigma$, the quantum relative entropy is given by
\begin{align}
    D(\rho||\sigma) = \Tr\rho(\ln\rho - \ln \sigma).
\end{align}
It is a non-negative $D(\rho||\sigma) \geq 0$ and non-symmetric measure of distinguishability. It vanishes if and only if $\rho = \sigma$ and it obeys the data processing inequality, $D(\mathcal{N}(\rho)||\mathcal{N}(\sigma)) \leq D(\rho||\sigma)$ for any quantum channel $\mathcal{N}(\cdot)$. The latter is an expression of the fact that physical operations cannot increase our ability to tell states apart.  

Now consider 2 logical spin states evolved under the channel $\mathcal{N}^{t}$
\begin{align}
    \rho_{S}(Q_{1}) &= \mathcal{N}^{t}(|\psi(Q_{1})\rangle \langle \psi(Q_{1})|) \nonumber \\
    \rho_{S}(Q_{2}) &= \mathcal{N}^{t}(|\psi(Q_{2})\rangle \langle \psi(Q_{2})|).
\end{align}
The relative entropy $D(\rho_{S}(Q_{1})||\rho_{S}(Q_{1}))$ quantifies the distinguishability of the two logical states at the final time $t$. Once again, averaging over the multiplicity tensors in the large-multiplicity limit yields
\begin{align}\label{eq:JLMS}
    \overline{D(\rho_{S}(Q_{1})||\rho_{S}(Q_{1}))} = D(\rho_{g} (Q_{1})||\rho_{g} (Q_{2}))
\end{align}
where $\rho_{g}(Q_{1})$ and $\rho_{g}(Q_{2})$ are decohered surface code logical states. 
\begin{align}
    \rho_{g}(Q_{1})  &= \mathcal{E}(|\Phi_{g} (Q_{1})\rangle \langle \Phi_{g} (Q_{1})|) \\
    \rho_{g}(Q_{2})  &= \mathcal{E}(|\Phi_{g} (Q_{2})\rangle \langle \Phi_{g} (Q_{2})|)
\end{align}
Since the surface-code logical states remain perfectly distinguishable at weak decoherence strength (in the $L\rightarrow \infty$ limit), so do the corresponding spin chain states. \eqnref{eq:JLMS} establishes an equivalence between bulk and boundary relative entropies; in the holography literature, this relation is known as the \textit{JLMS} formula~\cite{JLMS}.

\section{Random Bond Clock Model}\label{sec:clock_model}
Here we assess the effect of topological frustration in the random bond clock model along the Nishimori line. The key property of the Nishimori line is that the disorder distribution $P(\bf{s})$ is effectively proportional to the partition function itself, $\mathcal{Z}[\bf{s}]$. For concreteness, we focus on the clock-model Hamiltonian with a single nonzero coupling $\beta_{1} = \beta$ which plays the role of an inverse temperature:
\begin{align}
    H[\{\theta_{i}\};\textbf{s}^{Q}] &= -\beta \sum_{\langle i, j\rangle} \cos{\frac{2\pi \big(\theta_{i} - \theta_{j}-s^{Q}_{{ij}}\big)}{N}}
\end{align}
We additionally assume periodic boundary conditions in the spatial direction, although none of our results depend on this choice.

\subsection{Low Temp Limit}
When $\beta \ggg 1$, the $\mathbb{Z}_{N}$ clock model lies deep in the ordered phase. In this regime, the partition function $\mathcal{Z}[\textbf{s};Q]$ for quenched set of couplings $\textbf{s}$ is dominated by minimal energy saddle point configurations. Any bond configuration $\bf{s}$ that contains local frustrations necessarily produces a saddle point with strictly higher energy, leading to a partition function that is exponentially suppressed in $\beta$. Since the Nishimori condition implies $P(\bf{s})\sim \mathcal{Z}[\bf{s}]$, all locally frustrated bond configurations must therefore be exponentially unlikely. This is physically natural since in this limit the weak measurements become effectively projective, and the probability of observing charge defects should be negligible.The same argument applies to topological frustration. For a fixed, locally frustration-free bond configuration $\textbf{s}$, only one of $\textbf{s}^{Q}$ (\eqnref{eq:twisted_bonds}) can have vanishing topological frustration, namely $\hat{\gamma}(\textbf{s}^{Q})=0$. This is because $\hat{\gamma}(\textbf{s}^{Q})=\hat{\gamma}(\textbf{s}^{0}) + Q \mod N$. We denote this configuration by $Q'$ with $\hat{\gamma}(\textbf{s}^{Q'})=0$. Note that the specific value of $Q'$ is determined by the measurement outcomes and cannot be predicted a priori. All other twists $Q\neq Q'$ produce a configuration with nonzero topological frustration. In this case, $\mathcal{Z}[\textbf{s};Q]$ contains a domain wall extending along the circuit-time direction. Thus, the associated energy cost yields an exponential suppression of the partition function relative to the untwisted sector,
\begin{align}
    \mathcal{Z}[\textbf{s};Q] \approx e^{-c(Q-Q')t}\mathcal{Z}[\textbf{s};Q'],
\end{align}
with $c(0)=0$. We therefore conclude that the global charge sharpens to $Q'$ on a timescale $t_{\#} \sim \mathcal{O}(1)$.

\subsection{High Temp Phase}
When $\beta \lll 1$, the $\mathbb{Z}_{N}$ clock model lies deep in the disordered (high-temperature) phase. In this regime the partition function $\mathcal{Z}[\mathbf{s};Q]$ becomes a nearly uniform sum over all current-loop configurations. Since energetics is irrelevant at leading order, local frustrations in the bond variables $\mathbf{s}$ have negligible effect, and thus $\mathcal{Z}[\mathbf{s};Q]$ is expected to be approximately independent of the topological sector $Q$. This intuition can be made precise via the standard high-temperature expansion. This is given by 
\begin{align}
    \mathcal{Z}[\textbf{s};Q]  &\approx \sum_{\{ \theta_{p}\}}  \prod_{\langle i,j\rangle}\bigg[1 + \beta\cos{\frac{2\pi \big(\theta_{i} - \theta_{j}-s^{Q}_{ij}\big)}{N}}  \bigg] 
\end{align}
Thus the high temperature expansion is an expansion in sums of products of links where each link $\langle i,j\rangle$ contributes a factor $ \beta\cos{\frac{2\pi (\theta_{i} - \theta_{j}-s^{Q}_{ij})}{N}} $. We can sum over the spins $\theta_{i}$ using the identities
\begin{align}
    \sum_{\theta =0}^{N-1}\cos{\frac{2\pi \big(\theta - \phi\big)}{N}} &= 0 \nonumber\\
    \sum_{\theta =0}^{N-1}\cos{\frac{2\pi \big(\theta - \phi_{1}\big)}{N}}\cos{\frac{2\pi \big(\theta - \phi_{2}\big)}{N}} &=\frac{N}{2}\cos{\frac{2\pi \big(\phi_{1} - \phi_{2}\big)}{N}} \nonumber
\end{align}
Only terms in which each $\theta_i$ appears an even number of times survive, yielding a sum over loops. For a single closed loop $\gamma$ one obtains a contribution $\beta^{|\gamma|}\cos{\frac{2\pi s^{Q}_{\gamma}}{N}}$ where $s^{Q}_{\gamma} = \sum_{\langle i,j\rangle \in \gamma} s^{Q}_{ij}$. Note that for any contractible loop $\gamma$, $s^Q_{\gamma}$ is independent of $Q$. Hence, all such loops contribute identically to every $\mathcal{Z}[\mathbf{s};Q]$, and thus, they cancel out in the expression for $\lAngle |T_{\hat{\gamma}}|^{2} \rAngle$. The leading contributions to $\lAngle  |T_{\hat{\gamma}}|^{2} \rAngle$ stem from non-contractible loops that wind around the cylinder. These loops have length of order $L$ with a degeneracy proportional to $t$ (the cylinder height). Thus, for any two topological sectors $Q$ and $Q'$, we expect
\begin{align}
    \mathcal{Z}[\mathbf{s};Q]-\mathcal{Z}[\mathbf{s};Q']
\approx a(Q-Q') te^{-L\ln(1/\beta)}.
\end{align}
This difference is exponentially suppressed in the system size $L$. We therefore conclude that the charge-fuzzy phase is extremely stable at small $\beta$ and the charge-sharpening timescale grows exponentially with $L$:
\begin{align}
    t_{\#} \sim e^{L}.
\end{align}

\subsection{QLRO Phase}
Finally, for $N \geq 5$, the $\mathbb{Z}_{N}$ clock model exhibits an intermediate phase between the disordered and ordered phases, characterized by quasi–long-range order (QLRO). In this regime, the low-energy excitations are effectively gapless spin waves, and the infrared theory is well described by a Gaussian action. Accordingly, we model
\begin{align}
     \mathcal{Z}[\mathbf{s};Q] &= \int\mathcal{D}\vphi e^{-H[\vphi,\textbf{s}_{J}]}\\
    H[\vphi,\textbf{s}^{Q}] &= \frac{\rho_{s}}{2}\int d^{2}x(\nabla \vphi(x) - \textbf{s}^{Q}(x))^{2},
\end{align}
where $\varphi$ is a coarse-grained phase field that is effectively non-compact in this phase (vortices are suppressed), and $\mathbf{s}^{Q}$ is the coarse-grained field that encodes the phase disorder in the topological sector $Q$. $\rho_s$ is an effective spin stiffness. If we interpret $\mathbf{E} = -\nabla \varphi$ as an electric field and $\mathbf{s}^{Q}$ as a polarization field, the problem maps onto a two-dimensional electrostatics problem whereby we are solving for the electric field in the presence of a background distribution of dipole moments described by $\mathbf{s}^{Q}$. According to the Helmholtz decomposition theorem, any sufficiently smooth vector field on a 2D domain can be written as 
\begin{align}
     \textbf{s}^{Q}(x) =  \textbf{s}_{\text{curl-free}}(x) +  \textbf{s}_{\text{div-free}}(x) + \textbf{s}_{\text{harm}}^{Q}(x),
\end{align}
where 
\begin{align}
    \nabla \times\textbf{s}_{\text{curl-free}}(x)=0 &\implies  \textbf{s}_{\text{curl-free}}(x) = \nabla \chi \\
    \nabla \cdot\textbf{s}_{\text{div-free}}(x)=0  &\implies  \textbf{s}_{\text{div-free}}(x) = \nabla \times\textbf{r} \\
    \nabla \times\textbf{s}_{\text{harm}}^{Q} &=  \nabla \cdot\textbf{s}_{\text{harm}}^{Q}=0.
\end{align} 
The curl-free component is ``pure-gauge" and can be removed by shifting the phase field as $\varphi \rightarrow \varphi - \chi $. This is the frustration-free part of the disorder. In contrast, the divergence-free and harmonic components cannot be removed in this manner. The divergence-free component encodes local frustrations in the disorder variables. But in the QLRO (spin-wave), phase vortices are energetically suppressed and cannot compensate or screen the divergence-free part of $\mathbf{s}^{Q}$. Thus, the presence of any such frustration necessarily results in an increased energy cost. Along the Nishimori line, these disorder configurations are therefore probabilistically suppressed and the disorder is arranged so that $\mathbf{s}^{Q}$ contains no divergence-free component. 

Consequently, the only part of $\mathbf{s}^{Q}$ that remains relevant in the spin-wave description is the harmonic (``zero") mode $\mathbf{s}^{Q}_{\text{harm}}$, which captures the dependence of the free energy on the topological sector $Q$. The above discussion  implies that in the spin-wave phase, 
\begin{align}
    \int_{\mathcal{\text{Contractile }\gamma}} \textbf{s}^{Q}(x) &= 0 \\
     \int_{\mathcal{\text{Non-Contractile }\gamma}} \textbf{s}^{Q}(x) &=  \int_{\mathcal{\text{Non-Contractile }\gamma}} \textbf{s}^{0}(x) + Q.
\end{align}
where the second line corresponds to integration along a non-contractible loop winding once around the cylinder. The polarization field $\textbf{s}^{Q}$ necessarily contains a non-trivial harmonic (topological) zero mode for all but one value of $Q$. Let us denote this value by $Q'$. This then allows us to write 
\begin{align}
    \textbf{s}^{Q} = \frac{Q-Q'}{L}\textbf{e}_{x}
\end{align}
where ${Q-Q'}$ is now shifted to the symmetric integer range $\{-\lfloor N/2\rfloor,\ldots,\lfloor N/2\rfloor\}$. The presence of such a harmonic mode increases the energy of the configuration. Assuming that the spin-wave field $\varphi$ is itself periodic on the cylinder, we find that the additional energy cost is precisely given by 
\begin{align}
    \int_{0}^{t} dt' \int_{0}^{L}dx \frac{(Q-Q')^{2}}{L^{2}} = (Q-Q')^{2}\frac{t}{L}.
\end{align}
Thus, we conclude that in the QLRO phase
\begin{align}
    \mathcal{Z}[\mathbf{s};Q] \approx \mathcal{Z}[\mathbf{s};Q'] e^{-\frac{\rho_{s}}{2}\frac{t}{L}(Q-Q')^{2}}.
\end{align}
For $t\ll L$, this suppression factor is essentially unity, so all winding sectors contribute with nearly equal weight and the system remains in the charge-fuzzy phase. In contrast, when $t \sim L $, the suppression becomes appreciable; the nontrivial winding sectors are exponentially disfavored, and the system effectively undergoes charge sharpening. This results in a linear sharpening timescale 
\begin{align}
    t_{\#} \sim L.
\end{align}

\end{document}